\documentclass[12pt,letterpaper]{article}
\pdfoutput=1
\usepackage{jheppub}
\usepackage{amsfonts, amsthm}
\usepackage[english]{babel}
\usepackage[utf8]{inputenc}
\usepackage{slashed}
\usepackage{mathrsfs}
\usepackage{amssymb}
\usepackage{color}
\hypersetup{unicode}

\newcommand{\eq}{\begin{equation}}
\newcommand{\feq}{\end{equation}}
\newcommand{\eqn}{\begin{eqnarray}}
\newcommand{\feqn}{\end{eqnarray}}

\newcommand{\ma}[1]{\mbox{$\mathcal{#1}$}}

\newcommand{\masf}[1]{\mbox{$\mathsf{#1}$}}

\title{Duality invariance in Fayet-Iliopoulos gauged supergravity}

\author[a,b]{Sergio L.~Cacciatori,}
\author[b,c]{Dietmar Klemm}
\author[b,c]{and Marco Rabbiosi}

\affiliation[a]{Department of Science and High Technology, \\
Universit\`a dell’Insubria, \\
Via Valleggio 11, I-22100 Como, Italy.}
\affiliation[b]{INFN, Sezione di Milano, \\
Via Celoria 16, I-20133 Milano, Italy.}
\affiliation[c]{Dipartimento di Fisica, Universit\`a di Milano, \\
Via Celoria 16, I-20133 Milano, Italy.}

\emailAdd{sergio.cacciatori@uninsubria.it}
\emailAdd{dietmar.klemm@mi.infn.it}
\emailAdd{marco.rabbiosi@mi.infn.it}
\preprint{IFUM-1048-FT}

\abstract{We propose a geometric method to study the residual symmetries in
$N=2$, $d=4$ $\text{U}(1)$ Fayet-Iliopoulos (FI) gauged supergravity.
It essentially involves the stabilization of the symplectic vector of gauge couplings
(FI parameters) under the action of the U-duality symmetry of the ungauged theory.
In particular we are interested in those transformations that act non-trivially on the
solutions and produce scalar hair and dyonic black holes from a given seed. We illustrate the
procedure for finding this group in general and then show how it works in some specific models.
For the prepotential $F=-iX^0X^1$, we use our method to add one more parameter to the
rotating Chow-Comp\`ere solution, representing scalar hair.
}


\begin{document}
\maketitle
\flushbottom

\section{Introduction}

Duality transformations have played, and continue to play, an important role in fundamental
developments in string theory, supergravity, quantum field theory as well as in the physics of black holes.
Perhaps the most relevant example for this is the fact that the five known string theories are
actually all related by a web of dualities, and correspond just to perturbative expansions of a single
underlying theory about a distinct point in the moduli space of quantum vacua,
cf.~e.g.~\cite{Schwarz:1996bh} for a review. This web contains in particular weak/strong coupling
dualities, of which the celebrated AdS/CFT correspondence \cite{Aharony:1999ti} is another famous example. 

Duality transformations have been instrumental also in the construction of black hole solutions in
string theory. Typically one reduces a higher-dimensional theory (in presence of Killing directions)
to lower dimensions, in particular to $d=3$, where all vector fields can be dualized to become scalars.
One gets then three-dimensional gravity coupled to a nonlinear sigma model, and employs
the global symmetries of the latter to obtain new black holes from a given seed.
This technique was used by Cveti\v{c} and Youm \cite{Cvetic:1996xz} to construct the most general
rotating five-dimensional black hole solution to toroidally compactified heterotic string theory, specified
by 27 charges, two rotational parameters and the ADM mass. In a similar way, Chow and
Comp\`ere \cite{Chow:2014cca} obtained the most general asymptotically flat, stationary, rotating,
nonextremal, dyonic black hole of four-dimensional $N=2$ supergravity coupled to 3 vector multiplets
(the so-called stu model). It generates through U-dualities the most general asymptotically flat, stationary black hole of $N=8$ supergravity.

Note that this typical structure of getting, after a Kaluza-Klein reduction, three-dimensional gravity
coupled to a nonlinear sigma model, is also crucial to prove full integrability in some particular cases, cf.~e.g.~\cite{Belinsky:1971nt,Figueras:2009mc}.

When (part of the) global symmetries of some given supergravity theory are gauged, as it typically happens
in AdS supergravity, the sigma model target space isometries are generically broken
by the presence of a scalar potential, so that the powerful solution-generating techniques described
above seem to break down. An instructive example is the timelike dimensional reduction of
four-dimensional Einstein-Maxwell gravity down to three dimensions, which gives
Euclidean gravity coupled to an $\text{SU}(2,1)/\text{S}(\text{U}(1,1)\times\text{U}(1))$
sigma model \cite{Breitenlohner:1998cv,Breitenlohner:1987dg}. Adding a cosmological constant to the
Einstein-Maxwell theory leads to a scalar potential in three dimensions, that breaks three of the eight
$\text{SU}(2,1)$ generators, corresponding to the generalized Ehlers and the two Harrison transformations. This leaves merely a semidirect product of a one-dimensional Heisenberg group and a translation group
$\mathbb{R}^2$ as residual symmetry \cite{Klemm:2015uba}. Although in this concrete example
the surviving symmetries cannot be used to generate new solutions from known ones, they may
nevertheless be useful in more general settings.

The aim of this paper is thus to provide a systematical and thorough investigation of the residual
symmetries in $N=2$, $d=4$ $\text{U}(1)$ Fayet-Iliopoulos (FI) gauged supergravity, elaborating
on \cite{Halmagyi:2013uza}, where a particular stu model was considered.
To this end, we shall use a geometric method, whose underlying idea is
the following: The on-shell global symmetry group of the ungauged theory is called U-duality, and
consists of the isometries of the special K\"ahler non-linear sigma model that act linearly also on the
field strengths via the symplectic embedding \cite{Breitenlohner:1987dg}.
For purely electric gaugings, the scalar potential generically spoils this invariance, but allowing also for
dyonic gaugings one can recover the whole U-duality invariance, at the price of
changing the vector of gauge couplings and so the physical theory.
We will call this group $U_{\text{fi}}$, that stands for fake internal symmetry group, which acts
on a solution by mapping it to other solutions of other theories. 
Given $U_{\text{fi}}$, we fix a generic choice of the coupling constants $\mathcal G$.
The true internal symmetry group $U_{\text i}$ of the gauged supergravity theory is then $S_{\mathcal G}$,
the stabilizer of $\mathcal G$ under the action of $U_{\text{fi}}$\footnote{As we will see later, this
is true up to possible $\text{U}(1)$ factors.}.

The remainder of this paper is organized as follows: In the next section, we briefly review the theory
we are interested in, namely $N=2$, $d=4$ $\text{U}(1)$ FI-gauged supergravity, and explain more
in detail the general idea outlined above. In section \ref{sec:symm-4examples} we explicitely
determine the residual symmetry group for four different prepotentials that are frequently used,
but we stress that our method is general, and can be applied to arbitrary prepotentials
and extended to $N=4$ and $N=8$ gauged supergravity theories as well. After that,
in section \ref{sec:scalar-hair},
it is shown how to apply the residual symmetries to generate new black hole solutions from
a given seed in each of the four cases. In section \ref{sec:ext-hypers} we comment on a possible extension
of our work to include also gauged hypermultiplets. Section \ref{Conclusions} contains our conclusions
and some final remarks. Some supplementary material is deferred to two appendices.

\section{General strategy}

\subsection{$N=2$, $d=4$ FI-gauged supergravity}
\label{subsec:sugra}

The bosonic sector of $N=2$, $d=4$ supergravity coupled to $n_{\text V}$ vector multiplets consists of
the vierbein $e^a{}_\mu$, $n_{\text V}+1$ vector fields $A^\Lambda_\mu$ with
$\Lambda=0,\dots n_{\text V}$ (the graviphoton plus $n_{\text V}$ other fields from the vector multiplets),
and $n_{\text V}$ complex scalar fields $z^i$ ($i=1,\dots,n_{\text V}$). The latter parametrize an
$n_{\text V}$-dimensional special K\"ahler manifold, i.e., a K\"ahler-Hodge manifold, with K\"ahler
metric $g_{i\bar\jmath}(z,\bar z)$, which is the base of a symplectic bundle with the covariantly
holomorphic sections\footnote{We use the conventions of \cite{Andrianopoli:1996cm}.}
\begin{equation}
 \ma{V}=\left(\begin{array}{c}
                   L^\Lambda\\
                   M_\Lambda
                  \end{array}\right), \qquad 
                  D_{\bar \imath}\ma{V}\equiv\partial_{\bar \imath}\ma{V}
                  -\frac{1}{2}\left(\partial_{\bar \imath}\ma{K}\right)\ma{V}=0\,,
\end{equation}
where $\ma{K}$ is the K\"ahler potential. $\ma{V}$ obeys the constraint
\begin{equation}
 \left\langle\ma{V}|\ma{\bar V}\right\rangle\equiv
 \bar{L}^\Lambda M_\Lambda-L^\Lambda \bar{M}_\Lambda=-i\,. \label{eq:sympcond}
\end{equation}
Alternatively one can introduce the explicitly holomorphic sections of a different symplectic bundle,
\begin{equation}
 v \equiv e^{-\mathcal{K}/2}\ma{V}\equiv\left(\begin{array}{c}
						  X^\Lambda\\
						  F_\Lambda
						  \end{array}\right)\,.
\end{equation}
In appropriate symplectic frames it is possible to choose a homogeneous function $F(X)$ of second degree,
called prepotential, such that $F_\Lambda=\partial_\Lambda F$.
In terms of the sections $v$ the constraint (\ref{eq:sympcond}) becomes
\begin{equation}
\label{eq:sympcond2}
 \left\langle v|\bar{v}\right\rangle\equiv\bar{X}^\Lambda F_\Lambda-X^\Lambda{\bar{F}}_\Lambda=
 -i e^{-\mathcal{K}}.
\end{equation}
The couplings of the vector fields to the scalars are determined by the
$(n_{\text V}+1)\times(n_{\text V}+1)$ period matrix \ma{N}, defined by the relations
\begin{equation}
 M_\Lambda = \ma{N}_{\Lambda\Sigma}\, L^\Sigma\,, 
 \qquad D_{\bar\imath}\bar{M}_\Lambda=\ma{N}_{\Lambda\Sigma}\,D_{\bar \imath}\bar{L}^\Sigma\,.
\end{equation}
If the theory is defined in a frame in which a prepotential exists, \ma{N} can be obtained from
\begin{equation}
  \label{eq:period_matrix_prep}
  \ma{N}_{\Lambda\Sigma}=\bar{F}_{\Lambda\Sigma}
  + 2i\frac{(N_{\Lambda\Gamma}X^\Gamma)(N_{\Sigma\Delta}X^\Delta)}{X^\Omega N_{\Omega\Psi}X^\Psi}\,,
\end{equation}
where $F_{\Lambda\Sigma}=\partial_\Lambda\partial_\Sigma F$ and $N_{\Lambda\Sigma}\equiv\mathrm{Im}(F_{\Lambda\Sigma})$.
Introducing the matrix\footnote{We defined $R=\mathrm{Re}\,\ma N$ and
$I=\mathrm{Im}\,\ma N$.}
\begin{equation}
 \ma{M}=\left(\begin{array}{cc}
 I + 
 R  I^{-1}  R & \,\,-  R I ^{-1} \\
- I ^{-1}  R &  I ^{-1} \\
\end{array}\right), \label{def-M}
\end{equation}
we have the important relation between the symplectic sections and their derivatives,
\begin{equation}
\frac12 (\mathcal M - i\Omega) = \Omega\bar{\mathcal V}\mathcal V\Omega + \Omega D_i\mathcal V
g^{i\bar\jmath}D_{\bar\jmath}\bar{\mathcal V}\Omega\,, \label{eq:sympid}
\end{equation}
with
\begin{equation}
\Omega = \left(\begin{array}{cc} 0 & -1 \\ 1 & 0 \end{array}\right)\,.
\end{equation}
The bosonic Lagrangian reads
\begin{equation}
\label{eq:mainaction}
\sqrt{-g}^{-1}\!\mathscr{L} = \frac R2 - g_{i\bar\jmath}\,\partial_{\mu}z^i\partial^{\mu}
\bar z^{\bar\jmath} + \frac14 I_{\Lambda\Sigma}F^{\Lambda\mu\nu}F^{\Sigma}{}_{\mu\nu} 
+ \frac14 R_{\Lambda\Sigma}F^{\Lambda\mu\nu}\star\! F^{\Sigma}{}_{\mu\nu} - V(z,\bar z)\,.
\end{equation}
In the case of dyonic $\text{U}(1)$ FI-gauging, the scalar potential has the form \cite{Dall'Agata:2010gj}
\begin{equation}
\label{eq:scal_pot}
V = g^{i\bar\jmath}D_i{\mathcal L} D_{\bar\jmath}\bar{\mathcal L} - 3{\mathcal L}\bar{\mathcal L}\,,
\end{equation}
where $\mathcal L = \langle\mathcal G,\mathcal V\rangle$, and
$\mathcal G=(g^\Lambda,g_\Lambda)^t$ denotes the symplectic vector of gauge couplings (FI
parameters).

\subsection{\label{FSS}Fake internal symmetries, stabilization and solutions}

The kinetic part of (\ref{eq:mainaction}) corresponds to the action of the ungauged theory, whose on-shell
global symmetry group is called U-duality, consisting of the isometries of the non-linear sigma model 
that act linearly also on the field strengths via the symplectic embedding \cite{Breitenlohner:1987dg}.
For purely electric gaugings, the scalar potential generically spoils this invariance, but, as is clear from
\eqref{eq:scal_pot}, for dyonic gauging one recovers the whole U-duality invariance, at the price of
changing the vector of gauge couplings and so the physical theory.
We will call this group $U_{\text{fi}}$, that stands for fake internal symmetry group\footnote{When the
special K\"ahler manifold is symmetric we define the Lie algebra $\mathfrak{u}_{\text{fi}}$ of
$U_{\text{fi}}$ through the equations (\ref{eq:invP}). The corresponding definition for nonsymmetric
special K\"ahler manifolds requires more care.}.
The action of $U_{\text{fi}}$ on a solution is the mapping to other solutions of other theories, in the same way in which some elements of the symplectic group map solutions of theories with different prepotential
into each other \cite{Dall'Agata:2010gj}, cf.~e.g.~\eqref{changequad}, \eqref{changestu}.

Given $U_{\text{fi}}$, we fix a choice of the coupling constants $\mathcal G$ and,
at least at the beginning, we suppose that they are generic. 
We want to underline that for abelian dyonic gaugings, the Maxwell equations remain homogeneous
and so the action (\ref{eq:mainaction}) doesn't have topological terms \cite{deWit:2005ub}.

The true internal symmetry group $U_{\text i}$ of the gauged supergravity theory is $S_{\mathcal G}$,
the stabilizer of $\mathcal G$ under the action of $U_{\text{fi}}$, up to possible $\text{U}(1)$ factors. 
This is obvious from the definition of the stabilizer,
\begin{equation}
S_{\mathcal G} = \{g \in U_{\text{fi}}\,|\, g\mathcal G = \mathcal G\}\,,
\label{eq:stab}
\end {equation}
which means that we impose to stay in the same theory, and this restricts of course the group of
internal symmetries.

By acting with $S\in S_{\mathcal G}$ on a given seed solution
$(\ma V,\ma G,\ma F_{\mu\nu})$\footnote{Actually we should write
$(\ma V,\ma G,\ma F_{\mu\nu},g_{\mu\nu})$, but since $S_{\cal G}$ does not act on the metric, we shall
suppress the dependence on $g_{\mu\nu}$.}
of the equations of motion, we can generate another configuration via the map
\begin{equation}
(\ma V,\ma G,\ma F_{\mu \nu}) \mapsto (\tilde{\ma V},\tilde{\ma G},\tilde{\ma F}_{\mu\nu}) :=
(S\ma V,S\ma G,S\ma F_{\mu\nu}) = (S\ma V,\ma G,S\ma F_{\mu \nu} )\,. 
\label{eq:actionffake}
\end{equation}
The transformed fields solve the field equations by construction\footnote{As is clear from the formalism
introduced in \cite{Dall'Agata:2010gj}, the application of $S\in S_{\mathcal G}$ on a static solution of the
BPS flow preserves the same amount of supersymmetry as the original configuration. In the rotating case,
the same is true if one considers electric gaugings only \cite{Meessen:2012sr}.}.
In general, the scalars transform nonlinearly under the corresponding isometry, the field strengths are
rotated and the metric is functionally invariant.

Technically, in order to determine $S_{\mathcal G}$, it is simpler to work with the corresponding algebra
\begin{equation}
\mathfrak s_{\mathcal G} = \{a \in \mathfrak u _{\text{fi}} \, | \,a\mathcal G = 0\}\,.
\label{eq:stabalg}
\end {equation}
There are some cases in which $U_{\text i}$ strictly contains $S_{\mathcal G}$, and this depends on
some particular symmetric structures of the model under consideration. Typically, this happens because
the symmetry of the model allows to act with some symplectic matrices in a more general
way than (\ref{eq:actionffake}), leaving nevertheless the theory invariant.

\section{Stabilization and symmetries for some prepotentials}
\label{sec:symm-4examples}

Now we want to apply these techniques to some specific prepotentials. Each of them exhibits different
peculiar features related to the geometry of the underlying special K\"ahler manifold, namely to the
symplectic embedding of the isometry group of the non-linear sigma model (cf.~app.~\ref{symemb}).

\subsection{Prepotential $F=-iX^0X^1$}
\label{subsec:-iX^0X^1}

This prepotential encodes a particular special K\"ahler structure on the symmetric manifold
$\text{SU}(1,1)/\text{U}(1)$. The symplectic section is $\mathcal V = (X^0,X^1,-iX^1,-iX^0 )^t$, and we
fix the couplings in a completely electric frame, $\ma G = (0,0,g_0,g_1)^t$. The solution to
(\ref{eq:invP}) defines the algebra $\mathfrak u_{\text{fi}}$,
\[
b_1 t_1 + b_2 t_2 + b_3 t_3 + b_4 t_4 =\left(\begin{array}{cccc}
b_4 & 0 & b_1 & b_2   \\
0 & -b_4 & b_2 & b_3  \\
-b_3 & -b_2 & -b_4 & 0 \\
-b_2 & -b_1 & 0 & b_4 \\
\end{array}\right)\,,
\]
to be the U-duality $\text{su}(1,1)$ plus a $\text{u}(1)$, generated by $t_2$, which acts trivially on the
$z^i$, as we will see shortly. From the stability equation (\ref{eq:stabalg}) one finds that
$\mathfrak s_{\mathcal G}$ is generated by 
\begin{equation}
s = t_2 - \frac{g_1}{g_0} t_1 - \frac{g_0}{g_1} t_3\,,
\end{equation}
so that $S_{\mathcal G}\subseteq\text{U}(1,1)$ is the 1-parameter subgroup
\begin{equation}
S = e^{\beta s} = \left(\begin{array}{cccc}
\cos^2\!\beta &\frac{g_1}{g_0} \sin^2\!\beta & - \frac{g_1}{g_0}\cos\beta\sin\beta & \cos\beta
\sin\beta \\
\frac{g_0}{g_1}\sin^2\!\beta & \cos^2\!\beta & \cos\beta\sin\beta & - \frac{g_0}{g_1}\cos\beta
\sin\beta \\
\frac{g_0}{g_1}\sin\beta\cos\beta & -\cos\beta\sin\beta & \cos^2\!\beta & \frac{g_0}{g_1}\sin^2\!\beta \\
-\cos\beta\sin\beta & \frac{g_1}{g_0}\cos\beta\sin\beta & \frac{g_1}{g_0}\sin^2\!\beta & \cos^2\!\beta
\end{array}\right)\,. \label{matrix-S}
\end{equation}
On the other hand, the $\text{U}(1)$ generated by $t_2$ is given by
\begin{equation}
T_{\alpha} = e^{\alpha t_2}=\left(\begin{array}{cccc}
\cos\alpha & 0 & 0 & \sin\alpha \\
0 & \cos\alpha & \sin\alpha & 0 \\
0 & -\sin\alpha & \cos\alpha & 0 \\
-\sin\alpha & 0 & 0 & \cos\alpha \end{array}\right)\,,
\label{talpha}
\end{equation}
and it transforms the section $\mathcal V$ according to
\begin{equation}
T_{\alpha} \mathcal V = e^{-i\alpha}\mathcal V\,. \label{eq:iso}
\end{equation}
The projective special K\"ahler coordinates are thus insensible to its action. The matrix $\mathcal M$
defined in \eqref{def-M} transforms as
\begin{equation}
T_{\alpha}^t\mathcal M T_{\alpha} = \mathcal M\,.
\label{invM1}
\end{equation}
One can thus act with $T_{\alpha}$ on $\ma F_{\mu\nu}$ only, leaving the equations of motion still
invariant. $T_\alpha$ is an example for a `field rotation matrix' that is commonly used to generate
non-BPS solutions, a technique first introduced in \cite{Ceresole:2007wx,LopesCardoso:2007qid}
and subsequently applied to gauged supergravity in \cite{Klemm:2012vm,Gnecchi:2012kb}.
In conclusion, the internal symmetry group of this model is
$U_{\text i}=\text{U}(1)\times\text{U}(1)\supset S_{\mathcal G}$, with the two $\text{U}(1)$ factors
identified respectively with $S$ and $T_\alpha$.

\subsection{\label{simquad}Prepotential $F=\frac i4 X^\Lambda\eta_{\Lambda\Sigma}X^\Sigma$}

The prepotential $F=\frac i4 X^\Lambda\eta_{\Lambda\Sigma}X^\Sigma$, with
$\eta_{\Lambda\Sigma}=\text{diag}(-1,1,...,1)$, describes a special K\"ahler structure on the symmetric
manifolds $\text{SU}(1,n_{\text V})/(\text{U}(1)\times\text{SU}(n_{\text V}))$.
The symplectic section reads 
\begin{equation}
\mathcal V = (X^\Lambda, \frac i2\eta_{\Lambda\Sigma}X^\Sigma)^t\,.
\end{equation}
Due to the linearity of $\mathcal V$ in the coordinates $X^\Lambda$, one can easily construct the
one-parameter subgroup
\[
L_{\alpha} = \left(\begin{array}{cccc}
\cos\alpha & 0 & 2\sin\alpha & 0 \\
0 & I_{n_{\text V}}\cos\alpha & 0 & - 2I_{n_{\text V}}\sin\alpha \\
-\frac12\sin\alpha & 0 & \cos\alpha & 0 \\
0 & \frac12 I_{n_{\text V}}\sin\alpha & 0 &  I_{n_{\text V}}\cos\alpha
\end{array}\right) \label{lalpha}
\]
of $\text{Sp}(2n_{\text V}+2,\mathbb{R})$, under which the section $\mathcal V$ transforms as
\begin{equation}
L_{\alpha} \mathcal V = e^{-i\alpha} \mathcal V\,. \label{Lalphaact}
\end{equation}
Since
\begin{equation}
L_{\alpha}^t\mathcal M L_{\alpha} = \mathcal M\,, \label{invM2}
\end{equation}
we can add a new parameter to all the solutions of this model by acting with $L_{\alpha}$ on
$\ma F_{\mu\nu}$ only.\\
The stability equation is slightly more involved. Notice that the case with only one vector multiplet is symplectically equivalent to $F=-iX^0X^1$, and thus the results for $n_{\text V}=1$ can be obtained
from the previous subsection by an appropriate symplectic rotation, cf.~app.~\ref{symemb}.\\
Let us discuss the general case of $n_{\text V}=n$ vector multiplets. Eq.~\eqref{eq:invP} defining the
algebra $\mathfrak u_{\text{fi}}$ is equivalent to
\begin{equation}
Q^t = -\eta Q\eta\,, \qquad S = -\frac14\eta R\eta\,. \label{u1n-I}
\end{equation}
These equations define an embedding of $\text{U}(1,n)$ into $\text{Sp}(2n+2,\mathbb{R})$. To see this,
let $z=A+iB\in\mathfrak u(1,n)$. Then, $z^t\eta+\eta z=0$ implies
\begin{equation}
A^t = -\eta A \eta\,, \qquad B^t\eta = \eta B\,,
\end{equation}
so $\eta B$ is symmetric. This suggests an embedding
\begin{equation}
\iota_\alpha:\, \mathfrak u(1,n) \longrightarrow \mathfrak {sp}(2n+2,\mathbb{R})\,, \qquad
A+iB \longmapsto 
\begin{pmatrix}
A & \alpha B\eta \\ -\frac 1\alpha \eta B & -A^t
\end{pmatrix},
\label{eq:embedding}
\end{equation}
for any real $\alpha\neq 0$. This is indeed an injective Lie algebra morphism, and its image consists of
the elements of $\mathfrak{sp}(2n+2,\mathbb{R})$ which solve \eqref{eq:invP} with
$F_\Lambda=\frac i\alpha\eta_{\Lambda\Sigma} X^\Sigma$. In particular, \eqref{u1n-I} selects $\iota_2$.\\
A basis for $\mathfrak {u}(1,n)$ is given by the matrices 
\begin{align}
\{A_a\}_{a=1}^{n(n+1)/2}\,, \qquad \{iB_k\}_{k=0}^{n(n+3)/2}\,,
\end{align}
where $A_a$ are a basis for the space of $(n+1)\times (n+1)$ real matrices $A$ such that $\eta A$ is
antisymmetric, and $B_k$ generate the space of $(n+1)\times (n+1)$ real matrices $B$ such that $\eta B$
is symmetric, with $B_0=I$, the identity matrix. The embedding extends obviously to the group level
via the exponential map, and, in particular, notice that
\begin{eqnarray}
\exp (\alpha\iota_2(iB_0)) = L_\alpha\,.
\end{eqnarray}
Let us now consider the symmetry group $S_{\mathcal G}$. If we set
\begin{eqnarray}
\mathcal G=(\underline 0,\underline g)^t=(0,\vec 0, g_0,\vec g)^t\,,
\end{eqnarray}
with $\vec g=(g_1,\ldots,g_n)$, then we see that the invariance of $\mathcal G$ is defined by the equations 
\begin{equation}
A^t\underline{g} = 0\,, \qquad B\eta\underline{g} = 0\,,
\end{equation}
which define a maximal compact subgroup\footnote{To be precise, this is the subgroup
$\text{S}(\text{U}(1)\times\text{U}(n))$.} $\text{U}(n)$ of
$\text{U}(1,n)$. To see this, let us first put\footnote{We assume $\underline g$ to be timelike
future-directed, i.e., $\eta^{\Lambda\Sigma}g_\Lambda g_\Sigma<0$, $g_0>0$.}
\begin{eqnarray}
\hat g:=\sqrt{-\underline g^2}\,,
\end{eqnarray}
and define $\Lambda_{\underline g}\in\text{SO}(1,n)$ by
\begin{eqnarray}
(g_0,\vec g) = (\hat g,\vec 0)\Lambda_{\underline g}\,.
\end{eqnarray}
Thus, $A$ (or $\eta B^t$) has $\underline g$ in the cokernel if and only if
$\Lambda_{\underline g}A\Lambda^{-1}_{\underline g}$ (or
$\Lambda_{\underline g}\eta B^t\Lambda_{\underline g}^{-1}$) has $(\hat g,\vec 0)$ in the cokernel. From
this we immediately get that $\mathfrak s_{\mathcal G}$ is generated by the elements of
$\mathfrak u(1,n)$ of the form 
\begin{eqnarray}
z_{\underline g} = \Lambda_{\underline g}^{-1}z\Lambda_{\underline g}\,,
\end{eqnarray}
where $z\in\mathfrak u(1,n)$ has vanishing first row and first column. Thus,
$z_{\underline g}\in\text{U}(n)$.\\
This provides also a way to realize an explicit construction of the group elements of $S_{\mathcal G}$.
One can choose e.g.~a generalized Gell-Mann basis \cite{Bertini:2005rc} for $\mathfrak{su}(n)$, add the
identity matrix $I_n$ and then embed the basis into $\mathfrak{u}(1,n)$ by adding a first row and column
of zeros. If we call $\{z_I\}_{I=0}^{n^2-1}$ such a basis for the compact subalgebra $\mathfrak u(n)$ of
$\mathfrak{su}(1,n)$, then 
\begin{displaymath}
\{\iota_2(z_I)\}_{I=0}^{n^2-1}
\end{displaymath} 
is a basis for $\mathfrak s_{\mathcal G_0}$, where $\mathcal G_0\equiv(0,\vec 0,\hat g,\vec 0)$. Then
we can explicitly construct the group elements by means of the Euler construction of
$S_{\mathcal G_0}$\footnote{In a similar way one can use the Iwasawa construction to obtain
the whole group $U_{\text{fi}}$, whose compact part is just
$S_{\mathcal G}$ \cite{Cacciatori:2014vna}.}, as in \cite{Bertini:2005rc,Cacciatori:2012qi}. Finally we have
\begin{eqnarray}
S_{\mathcal G} = \tilde{\Lambda}_{\underline g}^{-1} S_{\mathcal G_0}\tilde{\Lambda}_{\underline g}\,,
\end{eqnarray}
with
\begin{equation}
\tilde{\Lambda}_{\underline g} =
\begin{pmatrix}
\Lambda_{\underline g} & 0 \\ 0 & \Lambda_{\underline g}^{-1}
\end{pmatrix}\,.
\end{equation}
For practical purposes we can take $\Lambda_{\underline g}$ defined by
\begin{equation}
{\Lambda_{\underline g}}^0{}_0 = \frac {g_0}{\hat g}\,, \qquad
{\Lambda_{\underline g}}^i{}_0 = {\Lambda_{\underline g}}^0{}_i = \frac {g_i}{\hat g}\,, \qquad
{\Lambda_{\underline g}}^i{}_j = \frac {g_0 - \hat g}{\hat g\vec g^2} g_i g_j + \delta^i{}_j\,,
\end{equation}
whose inverse is obtained by the replacement $\vec g\to -\vec g$.\\
Let us focus on the first nontrivial case $\text{SU}(1,2)/(\text{U}(1)\times\text{SU}(2))$. 
We fix the couplings in a completely electric frame, $\ma G = (0,0,0,g_0,g_1,g_2)^t$. A basis for
$\mathfrak u(2)$ (relative to the vector $\mathcal G_0=(0,\vec 0,\hat g,\vec 0)$) is
\begin{equation}
t_0 =
\begin{pmatrix}
0 & 0 & 0 \\ 0 & i & 0 \\ 0 & 0 & i
\end{pmatrix}\,, \quad
t_1 =
\begin{pmatrix}
0 & 0 & 0 \\ 0 & 0 & i \\ 0 & i & 0
\end{pmatrix}\,, \quad
t_2 = 
\begin{pmatrix}
0 & 0 & 0 \\ 0 & 0 & -1 \\ 0 & 1 & 0
\end{pmatrix}\,, \quad
t_3 =
\begin{pmatrix}
0 & 0 & 0 \\ 0 & i & 0 \\ 0 & 0 & -i
\end{pmatrix}\,,
\end{equation}
which, by means of $\iota_2$, defines the basis of $\mathfrak s_{\mathcal G_0}$
\begin{align}
T_0&= 
\begin{pmatrix}
0 & 0 & 0 & 0 & 0 & 0 \\ 0 & 0 & 0 & 0 & 2 & 0 \\ 0 & 0 & 0 & 0 & 0 & 2\\
0 & 0 & 0 & 0 & 0 & 0 \\ 0 & -\frac 12 & 0 & 0 & 0 & 0 \\ 0 & 0 & -\frac 12 &0 & 0 & 0
\end{pmatrix}\,,
&T_1= 
\begin{pmatrix}
0 & 0 & 0 & 0 & 0 & 0 \\ 0 & 0 & 0 & 0 & 0 & 2 \\ 0 & 0 & 0 & 0 & 2 & 0 \\
0 & 0 & 0 & 0 & 0 & 0 \\ 0 & 0 & -\frac 12 & 0 & 0 & 0 \\ 0 & -\frac 12 & 0 & 0 & 0 & 0
\end{pmatrix}\,,  \cr
T_2&= 
\begin{pmatrix}
0 & 0 & 0 & 0 & 0 & 0 \\ 0 & 0 & -1 & 0 & 0 & 0 \\ 0 & 1 & 0 & 0 & 0 & 0 \\
0 & 0 & 0 & 0 & 0 & 0 \\ 0 & 0 & 0 & 0 & 0 & -1 \\ 0 & 0 & 0 & 0 & 1 & 0
\end{pmatrix}\,,
&T_3=
\begin{pmatrix}
0 & 0 & 0 & 0 & 0 & 0 \\ 0 & 0 & 0 & 0 & 2 & 0 \\ 0 & 0 & 0 & 0 & 0 & -2 \\
0 & 0 & 0 & 0 & 0 & 0 \\ 0 & -\frac 12 & 0 & 0 & 0 & 0 \\ 0 & 0 & \frac 12 & 0 & 0 & 0
\end{pmatrix}\,.
\end{align}
Note that
\begin{displaymath}
T_0^2 = -\Delta\,, \qquad [T_i,T_j]_+ = -\delta_{ij}\Delta\,, \quad 1\leq i\leq j\leq 3\,, 
\end{displaymath}
with 
\begin{eqnarray}
\Delta=
\begin{pmatrix}
0 & 0 & 0 & 0 & 0 & 0 \\ 0 & 1 & 0 & 0 & 0 & 0 \\ 0 & 0 & 1 & 0 & 0 & 0 \\
0 & 0 & 0 & 0 & 0 & 0 \\ 0 & 0 & 0 & 0 & 1 & 0 \\ 0 & 0 & 0 & 0 & 0 & 1
\end{pmatrix}\,,
\end{eqnarray}
from which we immediately get the expression for a generic element of $S_{\mathcal G_0}$,
\begin{align}
S_0(x^0,\vec x)&=e^{x^0 T_0}e^{\vec x\cdot \vec T} \cr
&=(I_6-2\sin^2 \frac {x^0}2 \Delta +\sin x^0 T_0)(I_6-2\sin^2 \frac {|\vec x|}2 \Delta +
\sin |\vec x|\ \vec x \cdot \vec T)\,,
\end{align}
where $\vec x=(x^1, x^2, x^3)$, $|\vec x|=\sqrt{\vec x\cdot \vec x}$, $\vec T=(T_1, T_2, T_3)$ and
$\vec x \cdot \vec T=\sum_{i=1}^3 x^i T_i$.\\
Finally, after setting
\begin{equation}
T^g_\mu = \tilde{\Lambda}_{\underline g}^{-1} T_\mu\tilde{\Lambda}_{\underline g}\,, \quad
\mu=0,1,2,3\,, \qquad \Delta_{\underline g} = \tilde{\Lambda}_{\underline g}^{-1} \Delta
\tilde{\Lambda}_{\underline g}\,,
\end{equation}
we get for a generic element of $S_{\mathcal G}$
\begin{eqnarray}
S_{\underline g}(x^0,\vec x) &=& \tilde{\Lambda}_{\underline g}^{-1} S_0(x^0,\vec x)
\tilde{\Lambda}_{\underline g} \label{eq:stabqua} \\
&=& (I_6 - 2\sin^2\frac{x^0}2\Delta_{\underline g} + \sin x^0 T^g_0)(I_6 - 2\sin^2\frac{|\vec x|}2
\Delta_{\underline g} + \sin|\vec x|\ \vec x\cdot\vec T^g)\,. \nonumber
\end{eqnarray}
In order to have even more manageable expressions for the matrices, it may be convenient to change to
the basis $R_\mu$ defined by
\begin{displaymath}
R_0 = T^g_0\,, \quad
R_1 = \frac{g_1^2 - g_2^2}{g_1^2 + g_2^2} T^g_1 - \frac{2g_1g_2}{g_1^2 + g_2^2} T^g_3\,, \quad
R_2 = T^g_2\,, \quad
R_3 = \frac{g_1^2 - g_2^2}{g_1^2 + g_2^2} T^g_3 + \frac {2g_1g_2}{g_1^2 + g_2^2} T^g_1\,.
\end{displaymath}

\subsection{Prepotential $F=-X^1X^2X^3/X^0$}
\label{subsec:stu}

This prepotential describes a special K\"ahler structure on the symmetric manifold
$\left(\text{SU}(1,1)/\text{U}(1)\right)^3$, the well-known stu model. This is symplectically equivalent
to the model with $F=-2i(X^0X^1X^2X^3)^{1/2}$, for which supersymmetric black holes with
purely electric gaugings are known analytically \cite{Cacciatori:2009iz}. After a symplectic
transformation to $F=-X^1X^2X^3/X^0$, the electric gaugings considered in \cite{Cacciatori:2009iz}
become $\mathcal G = (0,g^1,g^2,g^3,g_0,0,0,0)^t$, so we shall concentrate on this case in what follows.
The symplectic section reads 
\begin{displaymath}
\mathcal V = (X^0, X^1, X^2, X^3, X^1 X^2 X^3/(X^0)^2, -X^2 X^3/X^0, -X^1 X^3/X^0,
-X^2 X^1/X^0 )^t\,.
\end{displaymath}
Let us now look at the solutions of \eqref{eq:invP}. To this end, we define
\begin{align}
\pmb X \equiv
\begin{pmatrix}
{X^0}^3 \\ {X^0}^2 X^1 \\ {X^0}^2 X^2 \\ {X^0}^2 X^3
\end{pmatrix}\,, \qquad
\pmb F \equiv
\begin{pmatrix}
X^1X^2X^3 \\ -X^0 X^2 X^3 \\ -X^0 X^1X^3 \\ -X^0 X^1 X^2
\end{pmatrix}\,,
\end{align}
so that \eqref{eq:invP} becomes
\begin{equation}
\pmb X S \pmb X -\pmb F R \pmb F -2 \pmb X Q^t \pmb F = 0\,. \label{eq:invP'}
\end{equation}
Since the lhs is a homogeneous polynomial of degree 6 in $(X^0,X^1,X^2,X^3)$, the coefficients of each
monomial must be zero. The simplest way to get the general solutions is then to look at the powers of
$X^0$. The possible powers of $X^0$ in $p_S\equiv\pmb X S\pmb X$, $p_R\equiv\pmb F R\pmb F$
and $p_Q\equiv\pmb X Q^t\pmb F$ are $(6,5,4)$, $(2,1,0)$ and $(4,3,2)$ respectively.
Since $S$ and $R$ are symmetric, $p_S$ and $p_R$ can vanish only if
$S$ and $R$ are zero. Thus, we are left with the following three possibilities:
\begin{enumerate}
\item $R=0$ and $p_Q$ cancels $p_S$. The only common power for $X^0$ is 4, so we have to take
matrices which generate only this power and equal degrees for the remaining variables. A quick inspection
gives the solutions\footnote{To avoid confusion, note that $S$ denotes the $4\times 4$ matrix
in \eqref{eq:invP'}, while $S_1$, $S_2$ and $S_3$ defined below are $8\times 8$ matrices.}
\begin{align}
S_1&=
\begin{pmatrix}
0 & 0 & 0 & 0 & 0 & 0 & 0 & 0 \\
0 & 0 & 0 & 0 & 0 & 0 & 0 & 0 \\
0 & 0 & 0 & 0 & 0 & 0 & 0 & 0 \\
-1 & 0 & 0 & 0 & 0 & 0 & 0 & 0 \\
0 & 0 & 0 & 0 & 0 & 0 & 0 & 1 \\
0 & 0 & 1 & 0 & 0 & 0 & 0 & 0 \\
0 & 1 & 0 & 0 & 0 & 0 & 0 & 0 \\
0 & 0 & 0 & 0 & 0 & 0 & 0 & 0 
\end{pmatrix}\,, \qquad
T_1=
\begin{pmatrix}
0 & 0 & 0 & 0 & 0 & 0 & 0 & 0 \\
0 & 0 & 0 & 0 & 0 & 0 & 0 & 0 \\
-1 & 0 & 0 & 0 & 0 & 0 & 0 & 0 \\
0 & 0 & 0 & 0 & 0 & 0 & 0 & 0 \\
0 & 0 & 0 & 0 & 0 & 0 & 1 & 0 \\
0 & 0 & 0 & 1 & 0 & 0 & 0 & 0 \\
0 & 0 & 0 & 0 & 0 & 0 & 0 & 0 \\
0 & 1 & 0 & 0 & 0 & 0 & 0 & 0 
\end{pmatrix}\,, \cr
U_1&=
\begin{pmatrix}
0 & 0 & 0 & 0 & 0 & 0 & 0 & 0 \\
-1 & 0 & 0 & 0 & 0 & 0 & 0 & 0 \\
0 & 0 & 0 & 0 & 0 & 0 & 0 & 0 \\
0 & 0 & 0 & 0 & 0 & 0 & 0 & 0 \\
0 & 0 & 0 & 0 & 0 & 1 & 0 & 0 \\
0 & 0 & 0 & 0 & 0 & 0 & 0 & 0 \\
0 & 0 & 0 & 1 & 0 & 0 & 0 & 0 \\
0 & 0 & 1 & 0 & 0 & 0 & 0 & 0 
\end{pmatrix}\,.
\end{align}
\item $S=0$ and $p_Q$ cancels $p_R$. The only common power for $X^0$ is 2, so we have to take
matrices generating only this and equal degrees for the remaining variables. The solution is
\begin{equation}
S_2 = S_1^t\,, \qquad T_2 = T_1^t\,, \qquad U_2 = U_1^t\,.
\end{equation}
\item $R=S=0$ and $Q$ satisfies $p_Q=0$. This implies that $Q$ must be diagonal and that the space
of such solutions is 3-dimensional. The simplest way to fix a basis of this space is to choose 
\begin{equation}
S_3 = [S_1,S_2]\,, \qquad T_3 = [T_1,T_2]\,, \qquad U_3 = [U_1,U_2]\,.
\end{equation}
\end{enumerate}
In this way the nine matrices $\vec S$, $\vec T$ and $\vec U$ generate the group
$U_{\text{fi}}=(\text{SL}(2,\mathbb{R}))^3$.\\
In order to determine the symmetry algebra $\mathfrak s_{\mathcal G}$ we have to consider the equation (using the same notation as in the previous subsection)
\begin{eqnarray}
(\vec x\cdot\vec S + \vec y\cdot\vec T + \vec z\cdot\vec U)\mathcal G = 0\,,
\end{eqnarray}
whose general solution is given by
\begin{displaymath}
{\cal U}(x,z) = g_0 g^3 x S_1 + g^1 g^2 x S_2 - g_0 g^2 (x+z) T_1 - g^1 g^3 (x+z) T_2 + g_0 g^1 z U_1
+ g^2 g^3 z U_2\,,
\end{displaymath}
for arbitrary $x,z\in\mathbb{R}$. A convenient basis is
\begin{equation}
{\cal U}_1 = {\cal U}(1,-1)\,, \qquad {\cal U}_2 = {\cal U}(1,0)\,, \label{eq:basis}
\end{equation}
which defines a two-dimensional abelian algebra. Notice that
\begin{equation}
 {\rm tr}\,{\cal U}_1^2 = {\rm tr}\,{\cal U}_2^2 = 8 g_0 g^1 g^2 g^3\,,
\end{equation}
so that the algebra is compact (and thus defines the group $\text{U}(1)\times\text{U}(1)$) if and only if
$g_0g^1g^2g^3<0$. One can easily verify that, unfortunately, none of these continuous symmetries
survives for the truncation to the $t^3$ model \cite{Fre:1995dw,Faedo:2015jqa} with prepotential
$F=-(X^1)^3/X^0$.\\
It is worth noting that a particular situation arises for $g^1=g^2=g^3=-g_0\equiv g$.  As was shown in
\cite{Halmagyi:2013uza}, there is an enhancement of the internal symmetry group in this case.
This happens because the scalar potential $V$ can be written in terms of fundamental objects that
define the nonlinear sigma model of the non-homogeneous projective coordinates $z^i=x^i+ iy^i$ \cite{Halmagyi:2013uza,Breitenlohner:1987dg}, namely
\begin{equation}
V = g^2\sum_{i=1}^3\mathrm{tr}M_i\,, \qquad
M_i = \left(\begin{array}{cc}
y^i + \frac{x^i{}^2}{y^i} & \frac{x^i}{y^i} \\
\frac{x^i}{y^i} & \frac1{y^i}
\end{array}\right)\,.
\end{equation}  
In fact, the transformation property of $M_i$,
\begin{equation}
M_i \longmapsto {\cal T}^t M_i\,{\cal T}\,,
\end{equation}
implies the invariance of the potential only if ${\cal T}{\cal T}^t=1$. Going back to the symplectic
formalism we see
that this condition is equivalent to require for the symmetry group to be orthogonal, which, in terms of
the elements of $\mathfrak u_{\text{fi}}$ amounts to consider just the subspace of antisymmetric matrices.
Thus, the symmetry algebra is generated by
\begin{equation}
W_1 = S_1 - S_2\,, \qquad W_2 = T_1 - T_2\,, \qquad W_3 = U_1 - U_2\,,
\label{enlarge}
\end{equation}
while the subalgebra leaving $\mathcal G$ fixed is generated by $W_2-W_1$ and $W_3-W_2$.
The full symmetry group is therefore an extension $U_{\text i}=\text{U}(1)^3$ of
$S_{\mathcal G}=\text{U}(1)^2$.

\subsection{Prepotential $F=X^1X^2X^3/X^0-\frac A3(X^3)^3/X^0$}

The base manifold for this prepotential is neither symmetric nor homogeneous and it has been studied
in \cite{Klemm:2015xda}. The symplectic section is given by $\mathcal V=(X^\Lambda,F_\Lambda)^t$, with
\begin{equation}
X^\Lambda{}^t = \begin{pmatrix} X^0 \\ X^1 \\ X^2 \\ X^3 \end{pmatrix}\,,
\qquad F_\Lambda^t = \begin{pmatrix} -X^1 X^2 X^3/(X^0)^2 + \frac A3(X^3)^3/(X^0)^2 \\
X^2 X^3/X^0 \\ X^1 X^3/X^0 \\ X^1 X^2/X^0 - A(X^3)^2/X^0 \end{pmatrix}\,.
\end{equation}
The solution to \eqref{eq:invP} is obtained by proceeding exactly like in the previous subsection. After
introducing the vectors
\begin{equation}
\pmb X=
\begin{pmatrix}
{X^0}^3 \\ {X^0}^2 X^1 \\ {X^0}^2 X^2 \\ {X^0}^2 X^3
\end{pmatrix}\,, \qquad
\pmb F=
\begin{pmatrix}
\frac A3 {X^3}^3-X^1X^2X^3 \\ X^0 X^2 X^3 \\ X^0 X^1X^3 \\ X^0 X^1 X^2-AX^0 {X^3}^2
\end{pmatrix}\,,
\end{equation}
we reduce the equations to a polynomial identity, and looking at the coefficients we get a five-dimensional
space of solutions generated by the symplectic matrices
\begin{displaymath}
S_1\! =\!
\begin{pmatrix}
0 & 0 & 0 & 0 & 0 & 0 & 0 & 0 \\
0 & 0 & 0 & 0 & 0 & 0 & 0 & 0 \\
0 & 0 & 0 & 0 & 0 & 0 & 0 & 0 \\
1 & 0 & 0 & 0 & 0 & 0 & 0 & 0 \\
0 & 0 & 0 & 0 & 0 & 0 & 0 & -1 \\
0 & 0 & 1 & 0 & 0 & 0 & 0 & 0 \\
0 & 1 & 0 & 0 & 0 & 0 & 0 & 0 \\
0 & 0 & 0 & -2A & 0 & 0 & 0 & 0 
\end{pmatrix}, \quad
S_2\! =\!
\begin{pmatrix}
0 & 0 & 0 & 0 & 0 & 0 & 0 & 0 \\
0 & 0 & 0 & 0 & 0 & 0 & 0 & 0 \\
1 & 0 & 0 & 0 & 0 & 0 & 0 & 0 \\
0 & 0 & 0 & 0 & 0 & 0 & 0 & 0 \\
0 & 0 & 0 & 0 & 0 & 0 & -1 & 0 \\
0 & 0 & 0 & 1 & 0 & 0 & 0 & 0 \\
0 & 0 & 0 & 0 & 0 & 0 & 0 & 0 \\
0 & 1 & 0 & 0 & 0 & 0 & 0 & 0 
\end{pmatrix}, \quad
S_3\! =\!
\begin{pmatrix}
0 & 0 & 0 & 0 & 0 & 0 & 0 & 0 \\
1 & 0 & 0 & 0 & 0 & 0 & 0 & 0 \\
0 & 0 & 0 & 0 & 0 & 0 & 0 & 0 \\
0 & 0 & 0 & 0 & 0 & 0 & 0 & 0 \\
0 & 0 & 0 & 0 & 0 & -1 & 0 & 0 \\
0 & 0 & 0 & 0 & 0 & 0 & 0 & 0 \\
0 & 0 & 0 & 1 & 0 & 0 & 0 & 0 \\
0 & 0 & 1 & 0 & 0 & 0 & 0 & 0 
\end{pmatrix},
\end{displaymath}
\begin{equation}
D_1 =
\begin{pmatrix}
3 & 0 & 0 & 0 & 0 & 0 & 0 & 0 \\
0 & 1 & 0 & 0 & 0 & 0 & 0 & 0 \\
0 & 0 & 1 & 0 & 0 & 0 & 0 & 0 \\
0 & 0 & 0 & 1 & 0 & 0 & 0 & 0 \\
0 & 0 & 0 & 0 & -3 & 0 & 0 & 0 \\
0 & 0 & 0 & 0 & 0 & -1 & 0 & 0 \\
0 & 0 & 0 & 0 & 0 & 0 & -1 & 0 \\
0 & 0 & 0 & 0 & 0 & 0 & 0 & -1 
\end{pmatrix}, \qquad
D_2 =
\begin{pmatrix}
0 & 0 & 0 & 0 & 0 & 0 & 0 & 0 \\
0 & 1 & 0 & 0 & 0 & 0 & 0 & 0 \\
0 & 0 & -1 & 0 & 0 & 0 & 0 & 0 \\
0 & 0 & 0 & 0 & 0 & 0 & 0 & 0 \\
0 & 0 & 0 & 0 & 0 & 0 & 0 & 0 \\
0 & 0 & 0 & 0 & 0 & -1 & 0 & 0 \\
0 & 0 & 0 & 0 & 0 & 0 & 1 & 0 \\
0 & 0 & 0 & 0 & 0 & 0 & 0 & 0 
\end{pmatrix}.
\end{equation}
A direct comparison with the results of \cite{Klemm:2015xda} shows that this algebra strictly contains the
U-duality algebra. This is due to the fact that the group of symmetries of the scalar potential is larger than
the symmetry group of the whole Lagrangian. Indeed the generator $D_2$ does not leave the metric
invariant. Thus, the U-duality group is generated by the algebra
\begin{eqnarray}
\langle S_1, S_2, S_3, D_1\rangle_{\mathbb R}\,.
\end{eqnarray}
Notice that the $S_i$ are nilpotent of order 4 for $i=1$ and order $2$ for $i=2,3$. They are indeed eigenmatrices for the adjoint action of $D_1$, all with eigenvalue $-2$.
The stability equation \eqref{eq:stabalg} has a nontrivial solution only if $A=-g^1g^2/(g^3)^2$.
With this choice for $A$ one gets a one-dimensional algebra $\mathfrak s_{\mathcal G} $ generated by
\begin{equation}
s = S_1 - \frac{g^1}{g^3} S_3 - \frac{g^2}{g^3} S_2\,. 
\label{nh}
\end{equation}
It is nilpotent of order $4$ so that $U_{\text i}=\mathcal S_{\mathcal G}$ is a unipotent group of order 4.
It is worthwhile to note that for $g^1=g^2=g^3$ one gets $A= -1$, which is the physically
most interesting case, since the corresponding prepotential arises in the context of type IIA string theory
compactifed on Calabi-Yau manifolds \cite{Louis:1996mt}.

\section{Scalar hair and dyonic solutions}
\label{sec:scalar-hair}

We shall now use the results of the previous section in order to generate new supergravity solutions
from a given seed. The transformations in $U_{\text i}$ add new parameters to a given solution
and leave not only the equations of motion invariant, but also some potential first-order flow
equations (if these are satisfied by the seed). The transformed field configuration preserves thus the
same amount of supersymmetry as the one from which we started.\\
As was stressed in \cite{Halmagyi:2013uza}, the latter statement is not true in the stu model for the
additional $\text{U}(1)$ that arises for equal couplings, whose action generically leads to a non-BPS
solution. The same story holds also in the quadratic models for $T_{\alpha}$ and $L_{\alpha}$, due to
the properties \eqref{invM1} and \eqref{invM2} \cite{Gnecchi:2012kb}.\\
In what follows we will consider several relevant examples for some well-studied prepotentials,
but there is no obstacle to extending this method to other solutions and prepotentials as well.
We underline that in the static case, owing to the existence of the black hole potential
$V_{\text{BH}}$ \cite{Ferrara:1995ih,Ferrara:1996dd}, one can directly rotate the charges $\mathcal Q$
instead of the field strengths $\mathcal F_{\mu\nu}$.

\subsection{Prepotential $F=-iX^0X^1$}

For this prepotential, we have $U_{\text i}=\text{U}(1)^2$, whose action on the static and magnetic BPS
seed solution of \cite{Cacciatori:2009iz} is
\begin{equation}
(\mathcal V, \mathcal G, \mathcal Q ) \quad \longmapsto \quad (\tilde{\mathcal V}, \tilde{\mathcal G},
\tilde{\mathcal Q}) = (S\mathcal V, \mathcal G, T_{\alpha}S\mathcal Q)\,.
\end{equation}
Using the results of section \ref{subsec:-iX^0X^1} and the constraints on the seed parameters
(cf.~\cite{Cacciatori:2009iz}), one gets
\begin{equation}
\begin{split}
&\tilde{\mathcal Q} = (p^0\cos\alpha, p^1\cos\alpha, -p^1\sin\alpha, -p^0\sin\alpha)^t\,, \\
&\tilde{z} = \frac{\tilde{X}^1}{\tilde{X}^0} = \frac{g_0}{g_1}\cdot\frac{g_1 z\cos\beta +
i g_0\sin\beta}{g_0\cos\beta + i g_1 z\sin\beta}\,, \qquad z \equiv \frac{X^1}{X^0}\,.
\end{split} \label{Qtzt-iX^0X^1}
\end{equation} 
The parameter $\beta$ does not modify the supersymmetry of the solution; for $\alpha=0$ the new
configuration satisfies again the BPS flow equations of \cite{Cacciatori:2009iz,Dall'Agata:2010gj}.
For $\alpha\neq 0$ one gets a solution that still obeys a first-order flow, but this time a non-BPS
one \cite{Gnecchi:2012kb}, driven by the fake superpotential
\begin{equation}
W = e^U|\langle T_{-\alpha}\tilde{\mathcal Q}, \tilde{\mathcal V}\rangle - i e^{2(\psi - U)}
\tilde{\mathcal L}|\,, \label{fake-W}
\end{equation}
where $U(r)$ and $\psi(r)$ are functions appearing in the metric
\begin{equation}
ds^2 = -e^{2U}dt^2 + e^{-2U}dr^2 + e^{2(\psi-U)}(d\theta^2 + \sinh^2\!\theta d\phi^2)\,,
\end{equation}
and $\mathcal L$ was defined in section \ref{subsec:sugra}. The first-order equations following
from \eqref{fake-W} imply the equations of motion provided the Dirac-type charge quantization condition
\begin{equation}
\langle\mathcal G, \mathcal Q\rangle = 1
\end{equation} 
holds \cite{Gnecchi:2012kb}. From \eqref{Qtzt-iX^0X^1} we see that for $\alpha\neq 0$ one
generates a dyonic solution from a purely magnetic one, while $\beta$ adds scalar hair to the seed.
Note that this result was first obtained in \cite{Halmagyi:2013uza}.

As another example for the action of $U_{\text i}$ we consider the Chow-Comp\`ere
solution \cite{Chow:2013gba}, that solves the equations of motion following from the Lagrangian
(2.12) of \cite{Chow:2013gba},
\begin{eqnarray}
\mathscr{L} &=& R\star\! 1 -\frac12\star\! d\varphi\wedge d\varphi -\frac12 e^{2\varphi}
\star\! d\chi\wedge d\chi - e^{-\varphi}\star\! F^1\wedge F^1 + \chi F^1\wedge F^1 \\
&& -\frac1{1+\chi^2 e^{2\varphi}}\left(e^{\varphi}\star\! F^2\wedge F^2 + \chi e^{2\varphi}F^2\wedge
F^2\right) + g^2\left(4 + e^{\varphi} + e^{-\varphi} + \chi^2 e^{\varphi}\right)\star\! 1\,, \nonumber
\end{eqnarray}
which is obtained from \eqref{eq:mainaction} by setting
\eq
z = \frac{g_0}{g_1}\left(e^{-\varphi} - i\chi\right)\,, \qquad g_0 g_1 = g^2\,,
\feq
and redefining\footnote{We assume $g_0/g_1>0$.}
\eq
F^0\longrightarrow\sqrt{\frac{g_1}{g_0}}F^1\,, \qquad F^1\longrightarrow\sqrt{\frac{g_0}{g_1}}F^2\,.
\feq
The dyonic rotating black hole solution of \cite{Chow:2013gba} is given by
\begin{equation}
ds^2 = -\frac RW\left(dt - \frac{a^2 - u_1u_2}a d\phi\right)^2 + \frac WR dr^2 + \frac UW\left(dt -
\frac{r_1r_2 + a^2}a d\phi\right)^2 + \frac WU du^2\,,
\end{equation}
where
\begin{eqnarray}
R(r) &=& r^2 - 2mr + a^2 + g^2 r_1 r_2(r_1 r_2 + a^2)\,, \nonumber \\
U(u) &=& -u^2 + 2nu + a^2 + g^2 u_1 u_2 (u_1 u_2 - a^2)\,, \\
W(r,u) &=& r_1 r_2 + u_1 u_2\,, \qquad r_{1,2} = r + \Delta r_{1,2}\,, \qquad u_{1,2} = u +
\Delta u_{1,2}\,, \nonumber
\end{eqnarray}
and $\Delta r_{1,2}$, $\Delta u_{1,2}$ are constants defined by
\begin{eqnarray}
\Delta r_1 &=& m[\cosh(2\delta_1)\cosh(2\gamma_2) - 1] + n\sinh(2\delta_1)\sinh(2\gamma_1)\,,
\nonumber \\
\Delta r_2 &=& m[\cosh(2\delta_2)\cosh(2\gamma_1) - 1] + n\sinh(2\delta_2)\sinh(2\gamma_2)\,,
\nonumber \\
\Delta u_1 &=& n[\cosh(2\delta_1)\cosh(2\gamma_2) - 1] - m\sinh(2\delta_1)\sinh(2\gamma_1)\,,
\nonumber \\
\Delta u_2 &=& n[\cosh(2\delta_2)\cosh(2\gamma_1) - 1] - m\sinh(2\delta_2)\sinh(2\gamma_2)\,.
\label{def-Delta-r-u}
\end{eqnarray}
Below we shall also use the linear combinations
\begin{eqnarray}
\Sigma_{\Delta r} &=& \frac12(\Delta r_1 + \Delta r_2)\,, \qquad \Delta_{\Delta r} = \frac12(\Delta r_2
- \Delta r_1)\,, \nonumber \\
\Sigma_{\Delta u} &=& \frac12(\Delta u_1 + \Delta u_2)\,, \qquad \Delta_{\Delta u} = \frac12(\Delta u_2
- \Delta u_1)\,. \label{def:Sigma-Delta}
\end{eqnarray}
The complex scalar field has the very simple form
\eq
z = \frac{g_0}{g_1}\frac{r_1 - i u_1}{r_2 - i u_2}\,,
\feq
while the gauge fields and their duals read
\begin{eqnarray}
A^1 &=& \zeta^1(dt - a d\phi) + \frac{r_2 u_2\tilde\zeta_1}{a}d\phi\,, \qquad
A^2 = \zeta^2(dt - a d\phi) + \frac{r_1 u_1\tilde\zeta_2}{a}d\phi\,, \nonumber \\
\tilde A_1 &=& \tilde\zeta_1(dt - a d\phi) - \frac{r_1 u_1\zeta^1}{a}d\phi\,, \qquad
\tilde A_2 = \tilde\zeta_2(dt - a d\phi) - \frac{r_2 u_2\zeta^2}{a}d\phi\,, \label{gauge-fields-CC}
\end{eqnarray}
where the three-dimensional electromagnetic scalars are
\begin{eqnarray}
\zeta^1 &=& \frac1{2W}\frac{\partial W}{\partial\delta_1} = \frac{Q_1r_2 - P^1u_2}W\,, \qquad
\tilde\zeta_1 = \frac{Q_1u_1 + P^1r_1}W\,, \nonumber \\
\zeta^2 &=& \frac1{2W}\frac{\partial W}{\partial\delta_2} = \frac{Q_2r_1 - P^2u_1}W\,, \qquad
\tilde\zeta_2 = \frac{Q_2u_2 + P^2r_2}W\,.
\end{eqnarray}
Here, $Q_{1,2}$ and $P^{1,2}$ denote respectively the electric and magnetic charges given
by \cite{Chow:2013gba}
\eq
Q_1 = \frac12\frac{\partial r_1}{\partial\delta_1}\,, \qquad Q_2 = \frac12\frac{\partial r_2}{\partial
\delta_2}\,, \qquad P^1 = -\frac12\frac{\partial u_1}{\partial\delta_1}\,, \qquad P^2 =
-\frac12\frac{\partial u_2}{\partial\delta_2}\,. \label{def-charges}
\feq
The solution is thus specified by the 7 parameters $m$, $n$, $a$, $\gamma_{1,2}$ and $\delta_{1,2}$
that are related to the mass, NUT charge, angular momentum, two electric and two magnetic charges.
Notice that a similar class of rotating black holes containing one parameter less was constructed
in \cite{Gnecchi:2013mja}.

Let us now consider the action of $S$ defined in \eqref{matrix-S}. For the transformed scalar we get
\begin{equation}
\tilde z = \frac{\tilde X^1}{\tilde X^0} = \frac{g_0}{g_1}\frac{r + \Delta r_1' - i(u + \Delta u_1')}{r +
\Delta r_2' - i(u + \Delta u_2')}\,,
\end{equation}
where
\begin{equation}
\left(\begin{array}{c} \Delta r_1' \\ \Delta r_2' \\ \Delta u_1' \\ \Delta u_2'\end{array}\right) =
\left(\begin{array}{cccc} \cos^2\!\beta & \sin^2\!\beta & -\cos\beta\sin\beta & \cos\beta\sin\beta \\
\sin^2\!\beta & \cos^2\!\beta & \cos\beta\sin\beta & -\cos\beta\sin\beta \\
\cos\beta\sin\beta & -\cos\beta\sin\beta & \cos^2\!\beta & \sin^2\!\beta \\
-\cos\beta\sin\beta & \cos\beta\sin\beta & \sin^2\!\beta & \cos^2\!\beta\end{array}\right)
\left(\begin{array}{c} \Delta r_1 \\ \Delta r_2 \\ \Delta u_1 \\ \Delta u_2\end{array}\right)\,.
\label{transf-Delta}
\end{equation}
Note that the quantities $\Sigma_{\Delta r}$ and $\Sigma_{\Delta u}$ defined in \eqref{def:Sigma-Delta}
remain invariant under \eqref{transf-Delta}, while $\Delta_{\Delta r}$ and $\Delta_{\Delta u}$
transform as
\begin{equation}
\left(\begin{array}{c} \Delta_{\Delta r}' \\ \Delta_{\Delta u}'\end{array}\right) = \left(\begin{array}{cc}
\cos2\beta & -\sin2\beta \\ \sin2\beta & \cos2\beta\end{array}\right)\left(\begin{array}{c}
\Delta_{\Delta r} \\ \Delta_{\Delta u}\end{array}\right)\,.
\end{equation}
The transformed gauge fields can be easily inferred from
\begin{equation}
\left(\begin{array}{c} A^1 + A^2 \\ \frac{g_1}{g_0}\tilde A_1 + \frac{g_0}{g_1}\tilde A_2 \\ A^2 - A^1 \\
\frac{g_0}{g_1}\tilde A_2 - \frac{g_1}{g_0}\tilde A_1\end{array}\right)^\prime = \left(\begin{array}{cccc}
1 & 0 & 0 & 0 \\ 0 & 1 & 0 & 0 \\ 0 & 0 & \cos2\beta & -\sin2\beta \\ 0 & 0 & \sin2\beta & \cos2\beta
\end{array}\right)\left(\begin{array}{c} A^1 + A^2 \\ \frac{g_1}{g_0}\tilde A_1 + \frac{g_0}{g_1}\tilde A_2 \\
A^2 - A^1 \\ \frac{g_0}{g_1}\tilde A_2 - \frac{g_1}{g_0}\tilde A_1\end{array}\right)\,.
\end{equation}
In conclusion, $S$ adds one more parameter $\beta$ to the solution of \cite{Chow:2013gba}.

Under the action of $T_\alpha$ (cf.~\eqref{talpha}) the scalar $z$ does not change. It turns out that the
new gauge fields can again be written in the form \eqref{gauge-fields-CC}, but with the three-dimensional
electromagnetic scalars replaced by
\begin{equation}
\left(\begin{array}{c} \sqrt\frac{g_1}{g_0}\zeta^1 \\ \sqrt\frac{g_0}{g_1}\zeta^2 \\
\sqrt\frac{g_1}{g_0}\tilde\zeta_1 \\ \sqrt\frac{g_0}{g_1}\tilde\zeta_2\end{array}\right)\longmapsto
\left(\begin{array}{cccc} \cos\alpha & 0 & 0 & \sin\alpha \\ 0 & \cos\alpha & \sin\alpha & 0 \\
0 & -\sin\alpha & \cos\alpha & 0 \\ -\sin\alpha & 0 & 0 & \cos\alpha\end{array}\right)
\left(\begin{array}{c} \sqrt\frac{g_1}{g_0}\zeta^1 \\ \sqrt\frac{g_0}{g_1}\zeta^2 \\
\sqrt\frac{g_1}{g_0}\tilde\zeta_1 \\ \sqrt\frac{g_0}{g_1}\tilde\zeta_2\end{array}\right)\,.
\end{equation}
In other words, they transform (up to prefactors) with the same matrix $T_\alpha$. This invariance can
be used to generate additional charges by starting from a given seed. Set e.g.~$\gamma_2=\delta_2=0$
in \eqref{def-Delta-r-u}, which by \eqref{def-charges} implies $P^2=Q_2=0$. After acting with
$T_\alpha$ one gets a solution with all four charges nonvanishing, namely
\begin{displaymath}
Q_1' = Q_1\cos\alpha\,, \qquad {P^1}' = P^1\cos\alpha\,, \qquad Q_2' = \frac{g_1}{g_0}P^1\sin\alpha\,,
\qquad {P^2}' = -\frac{g_1}{g_0}Q_1\sin\alpha\,.
\end{displaymath}

\subsection{Prepotential $F=\frac i4((X^1)^2+(X^2)^2-(X^0)^2)$}

In this case the most interesting feature of $U_{\text i}$ is the non-abelianity of
$\mathcal S_{\mathcal G}$, cf.~sec.~\ref{simquad}. As far as $L_{\alpha}$ is concerned, its effect
is the same as the one of $T_\alpha$ for $F=-iX^0X^1$, namely the transformed configuration
solves non-BPS first-order flow equations.\\
The nonabelian part acts nontrivially on the special scalars. With the 1-parameter subgroups
$\exp(\alpha_\mu R_\mu)$ ($\mu=0,\ldots,3$, no summation over $\mu$), where the $R_\mu$ are
defined in section \ref{simquad}, one can describe the action of $\mathcal S_{\mathcal G}$ on a static
seed solution with charge vector $\mathcal Q$ as
\begin{displaymath}
\begin{split}
&(\mathcal V, \mathcal G, \mathcal Q)\quad\longmapsto\quad (\tilde{\mathcal V},
\tilde{\mathcal G}, \tilde{\mathcal Q}) = (e^{\alpha_0 R_0}\mathcal V, \mathcal G,
e^{\alpha_0 R_0}\mathcal Q)\,, \\
&\tilde{z}^1=\frac{-g_1 (g_0 + g_1 z^1 + g_2 z^2) + e^{i\alpha_0} (g_0 g_1 +(g_0^2 - g_2^2) z^1 + g_1 g_2 z^2)}{g_0 (g_0 + g_1 z^1 + g_2 z^2) - e^{i\alpha_0}(g_1^2 + g_2^2 +g_0 g_2 z^2 + g_0 g_1 z^1)}\,, \\
&\tilde{z}^2=\frac{-g_2 (g_0 + g_1 z^1 + g_2 z^2) + e^{i\alpha_0} (g_0 g_2 +(g_0^2 - g_1^2) z^2 + g_1 g_2 z^1)}{g_0 (g_0 + g_1 z^1 + g_2 z^2) - e^{i\alpha_0} (g_1^2 + g_2^2 +g_0 g_2 z^2 + g_0 g_1 z^1)}\,,
\end{split}
\end{displaymath}

\begin{displaymath}
\begin{split}
&(\mathcal V, \mathcal G, \mathcal Q)\quad\longmapsto\quad (\tilde {\mathcal V},
\tilde{\mathcal G},\tilde{\mathcal Q}) = (e^{\alpha_1 R_1}\mathcal V, \mathcal G,
e^{\alpha_1 R_1}\mathcal Q)\,, \\
&\tilde{z}^1=\frac{-g_1 (g_0 +g_1 z^1 +g_2 z^2) + (g_0 g_1 +g_0^2 z^1 -g_2^2 z^1 + g_1 g_2 z^2) \cos \alpha_1 - \hat g (g_2 + g_0 z^2) \sin \alpha_1}{g_0(g_0 +g_1 z^1 + g_2 z^2) - (g_1^2 + g_0 g_1 z^1 + g_2^2 +g_0 g_2 z^2 )\cos\alpha_1 + \hat g (g_1 z^2 - g_2 z^1)\sin\alpha_1}\,, \\
&\tilde{z}^2=\frac{-g_2 (g_0 +g_1 z^1 +g_2 z^2) + (g_0 g_2 + g_0^2 z^2 -g_1^2 z^2 + g_2 g_1 z^1)
\cos \alpha_1 + \hat g (g_1 + g_0 z^1)\sin\alpha_1}{ g_0 (g_0 +g_1 z^1 + g_2 z^2) - (g_1^2 +
g_0 g_1 z^1 + g_2^2 +g_0 g_2 z^2 )\cos\alpha_1 + \hat g (g_1 z^2 - g_2 z^1)\sin\alpha_1}\,,
\end{split}
\end{displaymath}

\begin{displaymath}
\begin{split}
&(\mathcal V, \mathcal G, \mathcal Q)\quad\longmapsto\quad (\tilde{\mathcal V},
\tilde{\mathcal G}, \tilde{\mathcal Q}) = (e^{\alpha_2 R_2}\mathcal V, \mathcal G,
e^{\alpha_2 R_2}\mathcal Q)\,, \\
&\tilde{z}^1=\frac{-g_1 (g_0+g_1 z^1+g_2 z^2)+f(g_1,g_2,z^1,z^2)\cos\alpha_2 - h(g_1,g_2,z^1,z^2)
\sin \alpha_2}{g_0(g_0 +g_1 z^1 +g_2 z^2)-(g_1^2 + g_0 g_1 z^1 + g_2 (g_2 + g_0 z^2))\cos\alpha_2 +
i \hat g(g_2 z^1 - g_1 z^2)\sin\alpha_2}\,, \\
&\tilde{z}^2=\frac{-g_2 (g_0+g_1 z^1+g_2 z^2)+f(g_2,g_1,z^2,z^1)\cos\alpha_2 + h(g_2,g_1,z^2,z^1) 
\sin\alpha_2}{g_0(g_0+g_1 z^1+g_2 z^2)-(g_1^2 + g_0 g_1 z^1+g_2(g_2 + g_0 z^2))\cos\alpha_2 +
i \hat g(g_2 z^1 - g_1 z^2)\sin\alpha_2}\,, \\
\end{split}
\end{displaymath}

\begin{displaymath}
\begin{split}
&(\mathcal V, \mathcal G, \mathcal Q)\quad\longmapsto\quad (\tilde{\mathcal V},
\tilde{\mathcal G}, \tilde{\mathcal Q}) = (e^{\alpha_3 R_3}\mathcal V, \mathcal G,
e^{\alpha_3 R_3}\mathcal Q)\,, \\
&\tilde{z}^1 = \frac{-g_1(g_1^2+g_2^2)(g_0+g_1 z^1+g_2 z^2) + e^{i\alpha_3} k(g_1,g_2,z^1,z^2)+
e^{-i\alpha_3}g_2 \hat g^2 (g_2 z^1 - g_1 z^2)}{(g_1^2 + g_2^2)\left(g_0(g_0 +
g_1 z^1+g_2 z^2) - e^{i\alpha_3}(g_1^2+g_0 g_1 z^1 + g_2^2 + g_0 g_2 z^2 )\right)}\,, \\
&\tilde{z}^2=\frac{-g_2(g_1^2+g_2^2)(g_0+g_1 z^1+g_2 z^2) + e^{i\alpha_3} k(g_2,g_1,z^2,z^1) +
e^{-i\alpha_3}g_1 \hat g^2 (g_1 z^2 - g_2 z^1)}{(g_1^2 + g_2^2)\left(g_0(g_0+g_1 z^1+g_2 z^2)
- e^{i\alpha_3}(g_1^2+g_0 g_1 z^1 + g_2^2 + g_0 g_2 z^2 )\right) }\,, \\
\end{split}
\end{displaymath}
where we used the definitions
\begin{equation}
\begin{split} 
&\hat g = \sqrt{g_0^2 - g_1^2 - g_2^2}\,, \qquad f(g_1,g_2,z^1,z^2) = g_0 g_1 + g_0^2 z^1 +
g_1 g_2 z^2 - g_2^2 z^1\,, \\
&h(g_1,g_2,z^1,z^2) = \frac{i\hat g}{g_1^2 + g_2^2}(2 g_0 g_1 g_2 z^1 + g_1^2(g_2 - g_0 z^2) +
g_2^2(g_2 + g_0 z^2))\,, \\
&k(g_1,g_2,z^1,z^2) =  g_0 g_1 (g_1^2 + g_0 g_1 z^1 + g_2^2 + g_0 g_2 z^2)\,.
\end{split}
\end{equation}
The explicit expressions for $\tilde{\mathcal Q}$ are not particularly enlightening, so we don't report
them here. One may apply the above transformations to the static and magnetic BPS seed given by
eqns.~(3.100) and (3.101) of \cite{Cacciatori:2009iz} to generate dyonic and axionic solutions.\\
Note that the form of \eqref{eq:stabqua} splits the dependence of the group coordinates from the
couplings. Defining the section $\mathcal V_{\underline g}=({\pmb X}_{\underline g},
{\pmb F}_{\underline g})^t\equiv\tilde{\Lambda}_{\underline g}\mathcal V$,
the action of $\mathcal S_{\mathcal G}$ becomes
$\tilde{\mathcal V}_{\underline g}=S_0(x^0,\vec x)\mathcal V_{\underline g}$ that more explicitly reads
\begin{equation}
\tilde{\pmb X}_{\underline g} =
\begin{pmatrix}
X_{\underline g}^0 \\
e^{i x^0}\left( X_{\underline g}^1\cos|\vec x| + i ((x^1 + i x^2) X_{\underline g}^2 + i x^3
X_{\underline g}^1)\sin|\vec x|\right)\\
e^{i x^0}\left( X_{\underline g}^2\cos|\vec x| + i ((x^1 - i x^2) X_{\underline g}^1 - i x^3
X_{\underline g}^2)\sin|\vec x|\right)
\end{pmatrix}\,.
\end{equation}
This split is independent of the parametrization of the group and so one can also use that of \cite{Cacciatori:2012qi,Bertini:2005rc}.

\subsection{Prepotential $F=-X^1X^2X^3/X^0$}

This model is related to the one with $F=-2i(X^0X^1X^2X^3)^{1/2}$ by a symplectic rotation
with the matrix \eqref{changestu}. As a seed solution we shall thus take the static magnetic BPS black
holes given by eqns.~(3.31)-(3.34) of \cite{Cacciatori:2009iz}, transformed to $F=-X^1X^2X^3/X^0$.
In this new frame, the vectors of charges and couplings are respectively given by
\begin{equation}
\mathcal Q = (p^0,0,0,0,0,q_1,q_2,q_3)^t\,, \qquad \mathcal G = (0,g^1,g^2,g^3,g_0,0,0,0)^t\,.
\end{equation}
Assuming $g_0g^1g^2g^3<0$ and defining $A\equiv(-g_0g^1g^2g^3)^{1/2}$, the finite
transformations $\exp(\alpha_1 {\cal U}_1)$ and $\exp(\alpha_2 {\cal U}_2)$ generated
by \eqref{eq:basis} act as
\begin{equation}
\begin{split}
&(\mathcal V, \mathcal G, \mathcal Q )\quad\longmapsto\quad (\tilde {\mathcal V}, \tilde{\mathcal G},
\tilde {\mathcal Q}) = (e^{\alpha_1 {\cal U}_1}\mathcal V, \mathcal G, e^{\alpha_1 {\cal U}_1}
\mathcal Q)\,, \\
&\tilde{z}^1 = \frac{A z^1\cos(A\alpha_1) + g_0 g^1\sin(A\alpha_1)}{A\cos(A\alpha_1) + z^1 g^2 g^3
\sin(A\alpha_1)}\,, \\
&\tilde{z}^2 = z^2\,, \\
&\tilde{z}^3 = \frac{A z^3\cos(A\alpha_1) - g_0 g^3\sin(A\alpha_1)}{A\cos(A\alpha_1) - z^3 g^1 g^2
\sin(A\alpha_1)}\,,
\end{split} \label{eq:transf-V_1}
\end{equation}
\begin{equation}
\begin{split}
&(\mathcal V, \mathcal G, \mathcal Q )\quad\longmapsto\quad (\tilde{\mathcal V}, \tilde{\mathcal G},
\tilde{\mathcal Q}) = (e^{\alpha_2 {\cal U}_2}\mathcal V, \mathcal G, e^{\alpha_2 {\cal U}_2}
\mathcal Q)\,, \\
&\tilde{z}^1 = z^1\,, \\
&\tilde{z}^2 = \frac{A z^2\cos(A\alpha_2) + g_0 g^2\sin(A\alpha_2)}{A\cos(A\alpha_2) + z^2 g^1 g^3
\sin(A\alpha_2)}\,, \\
&\tilde{z}^3 = \frac{A z^3\cos(A\alpha_2) - g_0 g^3\sin(A\alpha_2)}{A\cos(A\alpha_2) - z^3 g^1 g^2
\sin(A\alpha_2)}\,.
\end{split} \label{eq:transf-V_2}
\end{equation}
Again, the expressions for $\tilde{\mathcal Q}$ are not particularly enlightening, so we shall not report
them here. Notice that the transformations \eqref{eq:transf-V_1}, \eqref{eq:transf-V_2} preserve the
supersymmetry of the seed.\\
As we pointed out in section \ref{subsec:stu}, in the special case $\mathcal G = (0,g,g,g,-g,0,0,0)^t$
there is an enhancement of the symmetry group to $\text{U}(1)^3$ generated by \eqref{enlarge}.
If we define $T=\exp[\frac{\alpha_3}3(W_1+W_2+W_3)]$, the action of the extra
$\text{U}(1)$ is
\begin{equation}
\begin{split}
&(\mathcal V, \mathcal G, \mathcal Q)\quad\longmapsto\quad (\tilde{\mathcal V}, \tilde{\mathcal G},
\tilde{\mathcal Q}) = (T\mathcal V, \mathcal G, T\mathcal Q )\,, \\
&\tilde{z}^1 = \frac{z^1\cos\alpha_3 - \sin\alpha_3}{z^1\sin\alpha_3 + \cos\alpha_3}\,, \\
&\tilde{z}^2 = \frac{z^2\cos\alpha_3 - \sin\alpha_3}{z^2\sin\alpha_3 + \cos\alpha_3}\,, \\
&\tilde{z}^3 = \frac{z^3\cos\alpha_3 - \sin\alpha_3}{z^3\sin\alpha_3 + \cos\alpha_3}\,,
\end{split} \label{eq:transf-T}
\end{equation}
plus an expression for the charges $\tilde{\mathcal Q}$. \eqref{eq:transf-V_1}, \eqref{eq:transf-V_2}
and \eqref{eq:transf-T} where first obtained in \cite{Halmagyi:2013uza}.
Note that $T$ breaks supersymmetry, since it
does not belong to the stabilizer $\mathcal S_{\mathcal G}$. In fact,
\begin{equation}
T\mathcal G\equiv\mathcal G_{\alpha_3} = g(\sin\alpha_3, \cos\alpha_3, \cos\alpha_3, \cos\alpha_3,
-\cos\alpha_3, \sin\alpha_3, \sin\alpha_3, \sin\alpha_3)^t\,.
\end{equation} 
However, the transformed solution still satisfies first-order non-BPS flow equations driven by the fake
superpotential \cite{Gnecchi:2012kb}\footnote{Notice that this flow is a BPS flow for a theory with
gaugings given by $\mathcal G_{\alpha_3}$.}
\begin{equation}
W = e^U|\langle\tilde Q, \tilde{\mathcal V}\rangle - i e^{2(\psi - U)}\langle\mathcal G_{\alpha_3},
\tilde{\mathcal V}\rangle|\,,
\end{equation}
provided the charge quantization condition $\langle\mathcal G,\mathcal Q\rangle=-\kappa$ holds,
where $\kappa=0,1,-1$ for flat, spherical or hyperbolic horizons respectively.

\subsection{Prepotential $F=X^1X^2X^3/X^0 +\frac{g^1g^2}{3(g^3)^2}(X^3)^3/X^0$}

In this case the only known solution with running scalars is that of \cite{Klemm:2015xda}, with static
metric and purely imaginary scalar fields,
\begin{equation}
X^1 /X^0 = z^1 = -i\lambda^1\,, \qquad X^2/X^0 = z^2 = -i\lambda^2\,, \qquad X^3/X^0 = z^3
= -i\lambda^3\,.
\end{equation}
The charges and coupling constants are given by
\begin{equation}
\mathcal Q = (p^0,0,0,0,0,q_1,q_2,q_3)^t\,, \qquad \mathcal G = (0,g^1,g^2,g^3,g_0,0,0,0)^t\,.
\end{equation}
Applying the finite transformation generated by \eqref{nh} yields for the scalars
\begin{equation}
\tilde z^1 = -i\lambda^1 - \frac{g^1}{g^3} c\,, \qquad\tilde z^2 = -i\lambda^2 - \frac{g^2}{g^3} c\,,
\qquad \tilde z^2 = -i\lambda^3 +c \,,
\end{equation}
and for the charges
\begin{equation}
\tilde{\mathcal Q}=\left(\begin{array}{c}
p^0 \\
-(c g^1 p^0)/g^3 \\
-(c g^2 p^0)/g^3 \\
c p^0 \\
- (4 c^3 g^1 g^2 p^0)/(3 {g^3}^2) +  (g^1 q_1 +g^2 q_2 -g^3 q_3)/g^3\\
q_1 - c^2 g^2 p^0/g^3 \\
q_2 - c^2 g^1 p^0/g^3 \\
q_3 +2 c^2 g^1 g^2 p^0/{g^3}^2
\end{array}
\right)\,,
\end{equation}
where $c$ is a group parameter. This solution is again BPS but has also nontrivial (constant) axions
turned on and all charges are nonvanishing.

\section{Extension to hypermultiplets}
\label{sec:ext-hypers}

In this section we briefly comment on a possible generalization of our work to include also hypermultiplets.
In this case the situation is more involved, since the coupling constants are replaced by the moment
maps $\ma P^x$. However, when only abelian isometries of the quaternionic hyperscalar target
space are gauged, the scalar potential can be cast into the form \cite{Klemm:2016wng}
\begin{equation}
V = \mathbb G^{AB}\mathbb D_A\ma L\,\mathbb D_B\bar{\ma L} - 3|\ma L|^2\,, 
\label{eq:potinv}
\end{equation}
where we defined
\begin{displaymath}
\mathbb G^{AB} = \left(\begin{array}{cc}
g^{i\bar\jmath} & 0 \\ 0 & h^{uv}\end{array}\right), \quad\mathbb D_A = \left(\begin{array}{l} D_i \\
\masf{D}_u\end{array}\right), \quad \ma L = \ma Q^x\ma W^x, \quad
\mathcal Q^x = \langle\ma P^x, \ma Q\rangle\,, \quad\ma W^x = \langle\ma P^x, \ma V\rangle\,.
\end{displaymath}
Here, $h_{uv}$ denotes the metric on the quaternionic manifold, and $\masf{D}_u$ is the covariant
derivative acting on the hyperscalars.\\
The most general symmetry transformation of the nonlinear sigma model
is a linear combination of the isometries of the quaternionic and the special K\"ahler manifold. Let us
define the formal operator
\begin{equation}
\delta = k^u\masf{D}_u + U \mathcal V \frac{\delta}{\delta\mathcal V} + U \bar{\mathcal V} \frac{\delta}{\delta\bar{\mathcal V}} 
+ U \mathcal A_{\mu} \frac{\delta}{\delta \mathcal A_{\mu}}
+ k^i\partial_i + k^{\bar\imath}\partial_{\bar\imath}\,,
\label{ew:general-vec}
\end{equation}
where $k^u$ is a Killing vector of the quaternionic manifold, $U$ an element of the U-duality algebra,
$k^i$ the corresponding holomorphic special K\"ahler Killing vector, and $ \mathcal A_{\mu} $ is 
the symplectic vector of the gauge potentials \cite{Klemm:2016wng}. Then it is clear from \eqref{eq:potinv}
that a sufficient condition for $\delta V=0$ is $\delta\mathcal L=0$\footnote{Note that, as in the FI
case, $\delta\mathcal L=0$ is in general sufficient but not necessary.}, that holds if and only
\begin{equation}
k^u\masf{D}_u\hat{\mathcal P}^x = U \hat{\mathcal P}^x\,, \label{eq:stab-hypers1}
\end{equation}
where we added a hat to the quaternionic quantities that define the gaugings.
Moreover the invariance of the kinetic term of the hyperscalars \cite{Andrianopoli:1996cm} leads to
\begin{equation}
({\cal L}_k \hat k)^v = U \hat{k}^v\,, \label{eq:stab-hypers2}
\end{equation}
where $\cal L$ denotes the Lie-derivative.
After choosing a specific model, these equations can in principle be solved for the parameters that define
the linear combination of Killing vectors (\ref{ew:general-vec}). In practice, (\ref{eq:stab-hypers1}) and
(\ref{eq:stab-hypers2}) represent a highly constrained and very model-dependent system, and it is a
priori not guaranteed that a nontrivial solution exists in general.
In the FI limit, \eqref{eq:stab-hypers1} boils down to the stabilization
equation for the coupling constants $\mathcal G$ and \eqref{eq:stab-hypers2} is trivially satisfied, as it
must be.\\
An interesting class of these models are the $N=2$ truncations of M-theory described 
in \cite{Cassani:2012pj,Halmagyi:2013sla}. In this case the solution of
(\ref{eq:stab-hypers1}) and (\ref{eq:stab-hypers2}) could simplify the study of the attractor equations \cite{Klemm:2016wng}, 
necessary to work along the lines of \cite{Benini:2015eyy}, namely to compare the gravity side 
with the recent field theory results of \cite{Hosseini:2016tor, Hosseini:2016ume, Benini:2016hjo}.

\section{\label{Conclusions}Conclusions}

In this paper we presented a geometric method to determine the residual symmetries in
$N=2$, $d=4$ $\text{U}(1)$ Fayet-Iliopoulos gauged supergravity.
It involves the stabilization of the symplectic vector of gauge couplings, i.e., the
FI parameters, under the action of the U-duality symmetry of the ungauged theory.
We then applied this to obtain the surviving symmetry group for a number of prepotentials frequently
used in the string theory literature, and showed how this group can be used to produce hairy and dyonic
black holes from a given seed solution. Moreover, we pointed out how our method may be extended
to a more general setting including also gauged hypermultiplets.

It would be very interesting to combine our results with dimensional reduction or oxidation
as a solution-generating technique much like in the ungauged case discussed in the introduction.
For instance one might think of starting from five-dimensional $N=2$ gauged supergravity coupled
to vector multiplets and then reduce to $d=4$ along a Killing direction to get one of the models
discussed here. One can then apply the residual symmetry group of the four-dimensional theory
and subsequently lift back to $d=5$ to generate new solutions. Notice that, for a timelike dimensional
reduction, the scalar manifold of the resulting Euclidean four-dimensional theory is para-K\"ahler
rather than K\"ahler \cite{Cortes:2009cs}, so that our results can not be applied straightforwardly, but
require some modifications. Another direction for future work could be to reduce gauged supergravity
theories to three dimensions and study in general the surviving symmetry preserved by the scalar potential.
Work along these directions is in progress \cite{CKR}.

\section*{Acknowledgements}

This work was supported partly by INFN. We would like to thank A.~Marrani, M.~Nozawa, N.~Petri and
A.~Santambrogio for useful discussions.

\appendix

\section*{\label{Appendix}Appendix}

\section{\label{Inv}Reparametrization and invariances}

A symplectic reparametrization of the section $\mathcal V$ for a prepotential $F=F(X)$ is a
transformation 
\begin{equation}
\mathcal V = (X^\Lambda, F_\Lambda)^t\longmapsto\tilde{\mathcal V} = (\tilde X^\Lambda,
\tilde F_\Lambda)^t\,.
\end{equation}
In the new frame a prepotential does not necessarily exist. We are interested in the subgroup of
$\text{Sp}(2n_{\text V}+2,\mathbb R)$ that leaves the prepotential
invariant \cite{deWit:1992wf,deWit:1991nm,Erbin:2014hsa},
\begin{equation}
F(\tilde{X}) = \tilde{F}(\tilde{X})\,.
\end{equation}
Its algebra is determined by the equation
\begin{equation}
X^\Lambda S_{\Lambda\Sigma} X^\Sigma - F_\Lambda R^{\Lambda\Sigma} F_\Sigma -2 X^\Lambda
Q^t{}_\Lambda{}^\Sigma F_\Sigma = 0\,, \label{eq:invP}
\end{equation}
where $Q$, $R$ and $S$ parametrize the symplectic algebra,
\[
U = \left(\begin{array}{cc} Q & R \\ S & -Q^t\end{array}\right)\,, \qquad R = R ^t\,, \qquad
S = S^t\,.
\]
A reparametrization of this type, in special projective coordinates, leaves $\ma V$ invariant up to a
K\"ahler transformation.

\section{\label{symemb}Symplectic embedding}

The choice of the symplectic embedding of the non-linear sigma model isometry group is necessary
to completely specify the special K\"ahler structure over a
manifold \cite{Andrianopoli:1996cm,Fre:1995dw,deWit:1991nm,deWit:1992wf,Cacciatori:2014vna}.
In what follows we shall summarize some properties used in the bulk of our paper.

\subsection{\label{equiemb}Symplectically equivalent embeddings}

The way in which the isometry group is embedded in the symplectic group is fixed by supersymmetry, and
in particular for $\text{SU}(1,n_{\text V})/(\text{U}(1)\times\text{SU}(n_{\text V}))$ and
$\text{SU}(1,1)/\text{U}(1)\times\text{SO}(2,2)/(\text{SO}(2)\times\text{SO}(2))$ one has
respectively \cite{Fre:1995dw}
\begin{equation}
(\mathbf{n_{\text V} + 1})\oplus (\mathbf{n_{\text V} +1})\qquad\mathrm{and}\qquad\mathbf 2\otimes
(\mathbf 4\oplus\mathbf 4)\,. \label{eq:emmb}
\end{equation}
This embedding is not unique since one can always act by conjugation with a symplectic matrix to
construct a symplectically equivalent embedding. There are choices for the section $\mathcal V$ such
that the isometry group sits in the symplectic group in a simple way, but the existence of a prepotential
in that frame is in general not guaranteed.
On the other hand, many symplectically equivalent embeddings are encoded by different prepotentials.
Two physically interesting examples are \cite{Sabra:1996xg,Gnecchi:2013mta}
\begin{equation}
\mathcal S_1 = \left(\begin{array}{cccc}
1 & 1 & 0 & 0 \\
1 & -1 & 0 & 0 \\
0 & 0 & \frac12 & \frac12 \\
0 & 0 & \frac12 & - \frac12
\end{array}\right)\,, \qquad -i X^0 X^1\longmapsto\frac i4 ({X^1}^2 - {X^0}^2)\,, \label{changequad}
\end{equation}
\begin{equation}
\mathcal S_2 = \left(\begin{array}{cccccccc}
1 & 0 & 0 & 0 & 0 & 0 & 0 & 0 \\
0 & 0 & 0 & 0 & 0 & 1 & 0 & 0  \\
0 & 0 & 0 & 0 & 0 & 0 & 1 & 0 \\
0 & 0 & 0 & 0 & 0 & 0 & 0 & 1 \\
0 & 0 & 0 & 0 & 1 & 0 & 0 & 0 \\
0 & -1 & 0 & 0 & 0 & 0 & 0 & 0  \\
0 & 0 & -1 & 0 & 0 & 0 & 0 & 0 \\
0 & 0 & 0 & -1 & 0 & 0 & 0 & 0
\end{array}\right)\,, \qquad -\frac{X^1 X^2 X^3}{X^0}\longmapsto -2i\sqrt{X^0 X^1 X^2 X^3}\,.
\label{changestu}
\end{equation}
A physically less important transformation, which is nevertheless useful for practical purposes, is for
instance
\begin{equation}
\mathcal S_a = \left(\begin{array}{cc}
a & 0 \\ 0 & \frac1a\end{array}\right)\,, \qquad\frac i4 X^\Lambda\eta_{\Lambda\Sigma} X^\Sigma
\longmapsto\frac i{4a^2} X^\Lambda\eta_{\Lambda\Sigma} X^\Sigma\,.
\end{equation}
One can also construct inequivalent embeddings over the same manifold, the simplest example being
$\text{SU}(1,1)/\text{U}(1)$ \cite{Fre:1995dw}. Notice finally that symplectic equivalence does not mean
physical equivalence. Even if it is possible to construct maps between the solutions of symplectically
equivalent models, in general the solutions are physically different.

\subsection{Special K\"ahler structure over $\text{SU}(1,n_{\text V})/(\text{U}(1)\times\text{SU}(n_{\text V}))$}

For this noncompact version of $\mathbb{C}\text{P}^n$ a simple way to embed
$\text{SU}(1,n_{\text V})$ into $\text{Sp}(2n_{\text V}+2,\mathbb R)$ is obtained from the fact that
\begin{equation}
\text{Sp}(2n_{\text V} + 2,\mathbb R)\cong\text{Usp}(1 + n_{\text V}, 1 + n_{\text V}) =
\text{Sp}(2n_{\text V} + 2,\mathbb C)\cap\text{U}(1 + n_{\text V}, 1 + n_{\text V})\,.
\end{equation}
This isomorphism is provided by conjugation with the Cayley matrix,
\begin{equation}
C_\alpha:\,\text{Sp}(2n_{\text V} + 2,\mathbb R)\longrightarrow\text{Usp}(1 + n_{\text V}, 1 +
n_{\text V})\,, \qquad U\longmapsto\hat{\mathcal C}_\alpha U\hat{\mathcal C}_\alpha^{-1}\,,
\end{equation}
where
\begin{equation}
\hat{\mathcal C}_\alpha = \frac1{\sqrt2}\left(\begin{array}{cc}
\frac1{\sqrt{\alpha}} I_{n_{\text V} + 1} & i\sqrt{\alpha}\eta \\
\frac1{\sqrt{\alpha}} I_{n_{\text V} + 1} & -i\sqrt{\alpha}\eta
\end{array}\right)\,,
\end{equation}
and $\eta$ is the Minkowski metric in $n_{\text V}+1$ dimensions. In fact
$\text{Usp}(1+n_{\text V},1+n_{\text V})$ is defined by the conditions
\begin{equation}
{\mathcal U}\mathbb H\,{\mathcal U}^\dagger = \mathbb H\,,\qquad {\mathcal U}\tilde{\Omega}
\,{\mathcal U}^t = \tilde{\Omega}\,. \label{def-Usp}
\end{equation}
If the invariant bilinear forms are chosen as
\begin{equation}
\mathbb H = \left(\begin{array}{cc} \eta & 0 \\ 0  & -\eta\end{array}\right)\,, \qquad
\tilde{\Omega} = \left(\begin{array}{cc} 0 & -\eta \\ \eta  & 0\end{array}\right)\,,
\end{equation}
\eqref{def-Usp} becomes
\begin{equation}
{\mathcal U} = \left(\begin{array}{cc} A & C^* \\ C & A^*\end{array}\right)\,, \qquad
A\eta A^\dagger - C^*\eta C^t = \eta\,, \qquad A^*\eta C^t - C\eta A^\dagger = 0\,. \label{simpleemb}
\end{equation}
The first of \eqref{eq:emmb} is obtained by restricting the action of $\iota_\alpha\equiv C_\alpha^{-1}$ to
the subgroup with $C=0$. One can also explicitly verify that in this frame the prepotential exists and is
given by $F=-\frac i{2\alpha}X^\Lambda\eta_{\Lambda\Sigma}X^\Sigma$.

\subsection{Special K\"ahler structure over
$\text{SU}(1,1)/\text{U}(1)\times\text{SO}(2,2)/(\text{SO}(2)\times\text{SO}(2))$}

This manifold belongs to the infinite sequence
$\text{SU}(1,1)/\text{U}(1)\times\text{SO}(2,n)/(\text{SO}(2)\times\text{SO}(n))$, which for $n=2$ is
isomorphic to $(\text{SL}(2,\mathbb R)/\text{SO}(2))^3$. To find the symplectic embedding it is useful
to choose a frame \cite{Fre:1995dw,Fre:1996js,Fre:2002pd,Fre':2003gd} in which the symplectic section
cannot be integrated to have a prepotential. In this frame the Calabi-Visentini parametrization appears in
a natural way. The embedding problem is solved by
\begin{equation}
\text{SO}(2,2)\ni L\longmapsto\left(\begin{array}{cc} L & 0 \\ 0 & {L^{-1}}^t\end{array}\right)\in
\text{Sp}(8,\mathbb R)\,,
\end{equation}
\begin{equation}
\text{SL}(2,\mathbb R)\ni\left(\begin{array}{cc} a & b \\ c & d\end{array}\right)\longmapsto
\left(\begin{array}{cc} a & b \hat\eta \\ c \hat\eta & d\end{array}\right)\in\text{Sp}(8,\mathbb R)\,,
\end{equation}
where $\hat\eta $ is the metric preserved by $\text{SO}(2,2)$. A symplectic transformation that leads to
a frame in which a prepotential exists is highly nontrivial to find \cite{Fre:1995dw}.


\begin{thebibliography}{99}

%\cite{Schwarz:1996bh}
\bibitem{Schwarz:1996bh}
  J.~H.~Schwarz,
  ``Lectures on superstring and M theory dualities: Given at ICTP spring school and at TASI summer school,''
  Nucl.\ Phys.\ Proc.\ Suppl.\  {\bf 55B} (1997) 1
  [hep-th/9607201].
  %%CITATION = doi:10.1016/S0920-5632(97)00070-4;%%

%\cite{Aharony:1999ti}
\bibitem{Aharony:1999ti}
  O.~Aharony, S.~S.~Gubser, J.~M.~Maldacena, H.~Ooguri and Y.~Oz,
  ``Large $N$ field theories, string theory and gravity,''
  Phys.\ Rept.\  {\bf 323} (2000) 183
  [hep-th/9905111].
  %%CITATION = doi:10.1016/S0370-1573(99)00083-6;%%

%\cite{Cvetic:1996xz}
\bibitem{Cvetic:1996xz}
  M.~Cveti\v{c} and D.~Youm,
  ``General rotating five-dimensional black holes of toroidally compactified heterotic string,''
  Nucl.\ Phys.\ B {\bf 476} (1996) 118
  [hep-th/9603100].
  %%CITATION = doi:10.1016/0550-3213(96)00355-0;%%

%\cite{Chow:2014cca}
\bibitem{Chow:2014cca}
  D.~D.~K.~Chow and G.~Comp\`ere,
  ``Black holes in $N=8$ supergravity from $\text{SO}(4,4)$ hidden symmetries,''
  Phys.\ Rev.\ D {\bf 90} (2014) no.2,  025029
  [arXiv:1404.2602 [hep-th]].
  %%CITATION = doi:10.1103/PhysRevD.90.025029;%%

%\cite{Belinsky:1971nt}
\bibitem{Belinsky:1971nt}
  V.~A.~Belinsky and V.~E.~Zakharov,
  ``Integration of the Einstein equations by the inverse scattering problem technique and the calculation
  of the exact soliton solutions,''
  Sov.\ Phys.\ JETP {\bf 48} (1978) 985
   [Zh.\ Eksp.\ Teor.\ Fiz.\  {\bf 75} (1978) 1953].
  %%CITATION = SPHJA,48,985;%%

%\cite{Figueras:2009mc}
\bibitem{Figueras:2009mc}
  P.~Figueras, E.~Jamsin, J.~V.~Rocha and A.~Virmani,
  ``Integrability of five-dimensional minimal supergravity and charged rotating black holes,''
  Class.\ Quant.\ Grav.\  {\bf 27} (2010) 135011
  [arXiv:0912.3199 [hep-th]].
  %%CITATION = ARXIV:0912.3199;%%

%\cite{Breitenlohner:1998cv}
\bibitem{Breitenlohner:1998cv}
  P.~Breitenlohner and D.~Maison,
  ``On nonlinear sigma models arising in (super-)gravity,''
  Commun.\ Math.\ Phys.\  {\bf 209} (2000) 785
  [gr-qc/9806002].
  %%CITATION = GR-QC/9806002;%%

%\cite{Breitenlohner:1987dg}
\bibitem{Breitenlohner:1987dg}
  P.~Breitenlohner, D.~Maison and G.~W.~Gibbons,
  ``Four-dimensional black holes from Kaluza-Klein theories,''
  Commun.\ Math.\ Phys.\  {\bf 120} (1988) 295.
  %%CITATION = doi:10.1007/BF01217967;%%

%\cite{Klemm:2015uba}
\bibitem{Klemm:2015uba}
  D.~Klemm, M.~Nozawa and M.~Rabbiosi,
  ``On the integrability of Einstein–Maxwell–(A)dS gravity in the presence of Killing vectors,''
  Class.\ Quant.\ Grav.\  {\bf 32} (2015) no.20,  205008
  [arXiv:1506.09017 [hep-th]].
  %%CITATION = doi:10.1088/0264-9381/32/20/205008;%%

%\cite{Halmagyi:2013uza}
\bibitem{Halmagyi:2013uza}
  N.~Halmagyi and T.~Vanel,
  ``AdS black holes from duality in gauged supergravity,''
  JHEP {\bf 1404} (2014) 130
  [arXiv:1312.5430 [hep-th]].
  %%CITATION = doi:10.1007/JHEP04(2014)130;%%

%\cite{Andrianopoli:1996cm}
\bibitem{Andrianopoli:1996cm}
  L.~Andrianopoli, M.~Bertolini, A.~Ceresole, R.~D'Auria, S.~Ferrara, P.~Fr\'e and T.~Magri,
  ``$N=2$ supergravity and $N=2$ super-Yang-Mills theory on general scalar manifolds: Symplectic
  covariance, gaugings and the momentum map,''
  J.\ Geom.\ Phys.\  {\bf 23} (1997) 111
  [hep-th/9605032].
  %%CITATION = doi:10.1016/S0393-0440(97)00002-8;%%

%\cite{Dall'Agata:2010gj}
\bibitem{Dall'Agata:2010gj}
  G.~Dall'Agata and A.~Gnecchi,
  ``Flow equations and attractors for black holes in $N=2$ $\text{U}(1)$ gauged supergravity,''
  JHEP {\bf 1103} (2011) 037
  [arXiv:1012.3756 [hep-th]].
  %%CITATION = doi:10.1007/JHEP03(2011)037;%%

%\cite{deWit:2005ub}
\bibitem{deWit:2005ub}
  B.~de Wit, H.~Samtleben and M.~Trigiante,
  ``Magnetic charges in local field theory,''
  JHEP {\bf 0509} (2005) 016
  [hep-th/0507289].
  %%CITATION = doi:10.1088/1126-6708/2005/09/016;%%

%\cite{Meessen:2012sr}
\bibitem{Meessen:2012sr}
  P.~Meessen and T.~Ort\'{\i}n,
  ``Supersymmetric solutions to gauged $N=2$ $d=4$ sugra: The full timelike shebang,''
  Nucl.\ Phys.\ B {\bf 863} (2012) 65
  [arXiv:1204.0493 [hep-th]].
  %%CITATION = doi:10.1016/j.nuclphysb.2012.05.023;%%

%\cite{Ceresole:2007wx}
\bibitem{Ceresole:2007wx}
  A.~Ceresole and G.~Dall'Agata,
  ``Flow equations for non-BPS extremal black holes,''
  JHEP {\bf 0703} (2007) 110
  [hep-th/0702088].
  %%CITATION = doi:10.1088/1126-6708/2007/03/110;%%

%\cite{LopesCardoso:2007qid}
\bibitem{LopesCardoso:2007qid}
  G.~Lopes Cardoso, A.~Ceresole, G.~Dall'Agata, J.~M.~Oberreuter and J.~Perz,
  ``First-order flow equations for extremal black holes in very special geometry,''
  JHEP {\bf 0710} (2007) 063
  [arXiv:0706.3373 [hep-th]].
  %%CITATION = doi:10.1088/1126-6708/2007/10/063;%%

%\cite{Klemm:2012vm}
\bibitem{Klemm:2012vm}
  D.~Klemm and O.~Vaughan,
  ``Nonextremal black holes in gauged supergravity and the real formulation of special geometry II,''
  Class.\ Quant.\ Grav.\  {\bf 30} (2013) 065003
  [arXiv:1211.1618 [hep-th]].
  %%CITATION = doi:10.1088/0264-9381/30/6/065003;%%

%\cite{Gnecchi:2012kb}
\bibitem{Gnecchi:2012kb}
  A.~Gnecchi and C.~Toldo,
  ``On the non-BPS first-order flow in $N=2$ $\text{U}(1)$-gauged supergravity,''
  JHEP {\bf 1303} (2013) 088
  [arXiv:1211.1966 [hep-th]].
  %%CITATION = doi:10.1007/JHEP03(2013)088;%%

%\cite{Bertini:2005rc}
\bibitem{Bertini:2005rc}
  S.~Bertini, S.~L.~Cacciatori and B.~L.~Cerchiai,
  ``On the Euler angles for $\text{SU}(N)$,''
  J.\ Math.\ Phys.\  {\bf 47} (2006) 043510
  [math-ph/0510075].
  %%CITATION = doi:10.1063/1.2190898;%%

%\cite{Cacciatori:2014vna}
\bibitem{Cacciatori:2014vna}
  S.~L.~Cacciatori, B.~L.~Cerchiai, S.~Ferrara and A.~Marrani,
  ``Iwasawa nilpotency degree of non compact symmetric cosets in $N$-extended supergravity,''
  Fortsch.\ Phys.\  {\bf 62} (2014) 350
  [arXiv:1402.5063 [hep-th]].

%\cite{Cacciatori:2012qi}
\bibitem{Cacciatori:2012qi}
  S.~L.~Cacciatori, F.~D.~Piazza and A.~Scotti,
  ``Compact Lie groups: Euler constructions and generalized Dyson conjecture,''
  arXiv:1207.1262 [math.GR].
  %%CITATION = ARXIV:1207.1262;%%

%\cite{Cacciatori:2009iz}
\bibitem{Cacciatori:2009iz}
  S.~L.~Cacciatori and D.~Klemm,
  ``Supersymmetric AdS$_4$ black holes and attractors,''
  JHEP {\bf 1001} (2010) 085
  [arXiv:0911.4926 [hep-th]].
  %%CITATION = doi:10.1007/JHEP01(2010)085;%%

%\cite{Fre:1995dw}
\bibitem{Fre:1995dw}
  P.~Fr\'e,
  ``Lectures on special K\"ahler geometry and electric-magnetic duality rotations,''
  Nucl.\ Phys.\ Proc.\ Suppl.\  {\bf 45BC} (1996) 59
  [hep-th/9512043].
  %%CITATION = doi:10.1016/0920-5632(95)00629-X;%%

%\cite{Faedo:2015jqa}
\bibitem{Faedo:2015jqa}
  F.~Faedo, D.~Klemm and M.~Nozawa,
  ``Hairy black holes in $N=2$ gauged supergravity,''
  JHEP {\bf 1511} (2015) 045
  [arXiv:1505.02986 [hep-th]].
  %%CITATION = doi:10.1007/JHEP11(2015)045;%%

%\cite{Klemm:2015xda}
\bibitem{Klemm:2015xda}
  D.~Klemm, A.~Marrani, N.~Petri and C.~Santoli,
  ``BPS black holes in a non-homogeneous deformation of the stu model of $N=2$, $D=4$ gauged     
  supergravity,''
  JHEP {\bf 1509} (2015) 205
  [arXiv:1507.05553 [hep-th]].
  %%CITATION = doi:10.1007/JHEP09(2015)205;%%

%\cite{Louis:1996mt}
\bibitem{Louis:1996mt}
  J.~Louis, J.~Sonnenschein, S.~Theisen and S.~Yankielowicz,
  ``Nonperturbative properties of heterotic string vacua compactified on $K3\times T^2$,''
  Nucl.\ Phys.\ B {\bf 480} (1996) 185
  [hep-th/9606049].
  %%CITATION = doi:10.1016/S0550-3213(96)00429-4;%%

%\cite{Ferrara:1995ih}
\bibitem{Ferrara:1995ih}
  S.~Ferrara, R.~Kallosh and A.~Strominger,
  ``$N=2$ extremal black holes,''
  Phys.\ Rev.\ D {\bf 52} (1995) 5412
  [hep-th/9508072].
  %%CITATION = doi:10.1103/PhysRevD.52.R5412;%%

%\cite{Ferrara:1996dd}
\bibitem{Ferrara:1996dd}
  S.~Ferrara and R.~Kallosh,
  ``Supersymmetry and attractors,''
  Phys.\ Rev.\ D {\bf 54} (1996) 1514
  [hep-th/9602136].
  %%CITATION = doi:10.1103/PhysRevD.54.1514;%%

%\cite{Chow:2013gba}
\bibitem{Chow:2013gba}
  D.~D.~K.~Chow and G.~Comp\`ere,
  ``Dyonic AdS black holes in maximal gauged supergravity,''
  Phys.\ Rev.\ D {\bf 89} (2014) no.6,  065003
  [arXiv:1311.1204 [hep-th]].
  %%CITATION = doi:10.1103/PhysRevD.89.065003;%%

%\cite{Gnecchi:2013mja}
\bibitem{Gnecchi:2013mja}
  A.~Gnecchi, K.~Hristov, D.~Klemm, C.~Toldo and O.~Vaughan,
  ``Rotating black holes in 4d gauged supergravity,''
  JHEP {\bf 1401} (2014) 127
  [arXiv:1311.1795 [hep-th]].
  %%CITATION = doi:10.1007/JHEP01(2014)127;%%

%\cite{Klemm:2016wng}
\bibitem{Klemm:2016wng}
  D.~Klemm, N.~Petri and M.~Rabbiosi,
  ``Symplectically invariant flow equations for $N=2$, $D=4$ gauged supergravity with hypermultiplets,''
  JHEP {\bf 1604} (2016) 008
  [arXiv:1602.01334 [hep-th]].
  %%CITATION = doi:10.1007/JHEP04(2016)008;%%

%\cite{Cassani:2012pj}
\bibitem{Cassani:2012pj}
  D.~Cassani, P.~Koerber and O.~Varela,
  ``All homogeneous $N=2$ M-theory truncations with supersymmetric AdS$_4$ vacua,''
  JHEP {\bf 1211} (2012) 173
  [arXiv:1208.1262 [hep-th]].
  %%CITATION = doi:10.1007/JHEP11(2012)173;%%

%\cite{Halmagyi:2013sla}
\bibitem{Halmagyi:2013sla}
  N.~Halmagyi, M.~Petrini and A.~Zaffaroni,
  ``BPS black holes in AdS$_4$ from M-theory,''
  JHEP {\bf 1308} (2013) 124
  [arXiv:1305.0730 [hep-th]].
  %%CITATION = doi:10.1007/JHEP08(2013)124;%%

%\cite{Benini:2015eyy}
\bibitem{Benini:2015eyy}
  F.~Benini, K.~Hristov and A.~Zaffaroni,
  ``Black hole microstates in AdS$_4$ from supersymmetric localization,''
  arXiv:1511.04085 [hep-th].
  %%CITATION = ARXIV:1511.04085;%%

%\cite{Hosseini:2016tor} 
\bibitem{Hosseini:2016tor}
  S.~M.~Hosseini and A.~Zaffaroni,
  ``Large $N$ matrix models for 3d ${\cal N}=2$ theories: twisted index, free energy and black holes,''
  arXiv:1604.03122 [hep-th].
  %%CITATION = ARXIV:1604.03122;%%

%\cite{Hosseini:2016ume}
\bibitem{Hosseini:2016ume}
  S.~M.~Hosseini and N.~Mekareeya,
  ``Large $N$ topologically twisted index: necklace quivers, dualities, and Sasaki-Einstein spaces,''
  arXiv:1604.03397 [hep-th].
  %%CITATION = ARXIV:1604.03397;%%

%\cite{Benini:2016hjo}
\bibitem{Benini:2016hjo}
  F.~Benini and A.~Zaffaroni,
  ``Supersymmetric partition functions on Riemann surfaces,''
  arXiv:1605.06120 [hep-th].
  %%CITATION = ARXIV:1605.06120;%%

%\cite{Cortes:2009cs}
\bibitem{Cortes:2009cs}
  V.~Cortes and T.~Mohaupt,
  ``Special geometry of Euclidean supersymmetry III: The Local r-map, instantons and black holes,''
  JHEP {\bf 0907} (2009) 066
  [arXiv:0905.2844 [hep-th]].
  %%CITATION = doi:10.1088/1126-6708/2009/07/066;%%

\bibitem{CKR}
  S.~L.~Cacciatori, D.~Klemm and M.~Rabbiosi, in preparation.

%\cite{deWit:1992wf}
\bibitem{deWit:1992wf}
  B.~de Wit, F.~Vanderseypen and A.~Van Proeyen,
  ``Symmetry structure of special geometries,''
  Nucl.\ Phys.\ B {\bf 400} (1993) 463
  [hep-th/9210068].
  %%CITATION = doi:10.1016/0550-3213(93)90413-J;%%

%\cite{deWit:1991nm}
\bibitem{deWit:1991nm}
  B.~de Wit and A.~Van Proeyen,
  ``Special geometry, cubic polynomials and homogeneous quaternionic spaces,''
  Commun.\ Math.\ Phys.\  {\bf 149} (1992) 307
  [hep-th/9112027].
  %%CITATION = doi:10.1007/BF02097627;%%

%\cite{Erbin:2014hsa}
\bibitem{Erbin:2014hsa}
  H.~Erbin and N.~Halmagyi,
  ``Abelian hypermultiplet gaugings and BPS vacua in $N=2$ supergravity,''
  JHEP {\bf 1505} (2015) 122
  [arXiv:1409.6310 [hep-th]].
  %%CITATION = doi:10.1007/JHEP05(2015)122;%%

%\cite{Sabra:1996xg}
\bibitem{Sabra:1996xg}
  W.~A.~Sabra,
  ``Symplectic embeddings and special K\"ahler geometry of $\mathbb{C}\text{P}(n-1,1)$,''
  Nucl.\ Phys.\ B {\bf 486} (1997) 629
  [hep-th/9608106].
  %%CITATION = doi:10.1016/S0550-3213(96)00697-9;%%

%\cite{Gnecchi:2013mta}
\bibitem{Gnecchi:2013mta}
  A.~Gnecchi and N.~Halmagyi,
  ``Supersymmetric black holes in AdS$_4$ from very special geometry,''
  JHEP {\bf 1404} (2014) 173
  [arXiv:1312.2766 [hep-th]].
  %%CITATION = doi:10.1007/JHEP04(2014)173;%%

%\cite{Fre:1996js}
\bibitem{Fre:1996js}
  P.~Fr\'e, L.~Girardello, I.~Pesando and M.~Trigiante,
  ``Spontaneous $N=2\rightarrow N=1$ local supersymmetry breaking with surviving compact gauge
  group,''
  Nucl.\ Phys.\ B {\bf 493} (1997) 231
  [hep-th/9607032].
  %%CITATION = doi:10.1016/S0550-3213(97)00076-X;%%

%\cite{Fre:2002pd}
\bibitem{Fre:2002pd}
  P.~Fr\'e, M.~Trigiante and A.~Van Proeyen,
  ``Stable de~Sitter vacua from $N=2$ supergravity,''
  Class.\ Quant.\ Grav.\  {\bf 19} (2002) 4167
  [hep-th/0205119].
  %%CITATION = doi:10.1088/0264-9381/19/15/319;%%

%\cite{Fre':2003gd}
\bibitem{Fre':2003gd}
  P.~Fr\'e, M.~Trigiante and A.~Van Proeyen,
  ``$N=2$ supergravity models with stable de Sitter vacua,''
  Class.\ Quant.\ Grav.\  {\bf 20} (2003) S487
  [hep-th/0301024].
  %%CITATION = doi:10.1088/0264-9381/20/12/314;%%






\end{thebibliography}
\end{document}